\documentclass[%
superscriptaddress,
reprint,
nofootinbib,
nobibnotes,
 amsmath,amssymb,
 aps,
]{revtex4-1}

\usepackage{graphicx}
\graphicspath{{Figures/}}
\usepackage{overpic}
\usepackage{morefloats}
\usepackage{dcolumn}
\usepackage{bm}

\usepackage[usenames,dvipsnames]{color}
\usepackage[normalem]{ulem}

\begin{document}

\title{Two-dimensional localized states in an active phase-field-crystal model}

\author{Lukas Ophaus}
\affiliation{Institut f\"ur Theoretische Physik, Westf\"alische
Wilhelms-Universit\"at M\"unster, Wilhelm-Klemm-Strasse 9, 48149
M\"unster, Germany}
 \affiliation{Center of Nonlinear Science (CeNoS), Westf\"alische
Wilhelms-Universit\"at M\"unster, Corrensstrasse 2, 48149 M\"unster,  Germany}

\author{Edgar Knobloch}
\affiliation{Department of Physics, University of California, Berkeley, California 94720, USA}

\author{Svetlana V. Gurevich}
\thanks{ORCID: 0000-0002-5101-4686}
\affiliation{Institut f\"ur Theoretische Physik, Westf\"alische
Wilhelms-Universit\"at M\"unster, Wilhelm-Klemm-Strasse 9, 48149
M\"unster, Germany}
 \affiliation{Center of Nonlinear Science (CeNoS), Westf\"alische
Wilhelms-Universit\"at M\"unster, Corrensstrasse 2, 48149 M\"unster,  Germany}

\author{Uwe Thiele}
 \thanks{ORCID: 0000-0001-7989-9271}
\email{u.thiele@uni-muenster.de}
\homepage{http://www.uwethiele.de}
 \affiliation{Institut f\"ur Theoretische Physik, Westf\"alische
Wilhelms-Universit\"at M\"unster, Wilhelm-Klemm-Strasse 9, 48149
M\"unster, Germany}
 \affiliation{Center of Nonlinear Science (CeNoS), Westf\"alische
Wilhelms-Universit\"at M\"unster, Corrensstrasse 2, 48149 M\"unster,  Germany}

\date{\today}

\begin{abstract}
  The active phase-field-crystal (active PFC) model provides a simple microscopic mean field description of crystallization in active systems. It combines the PFC model (or conserved Swift-Hohenberg equation) of colloidal crystallization and aspects of the Toner-Tu theory for self-propelled particles. We employ the active PFC model to study the occurrence of localized and periodic active crystals in two spatial dimensions. Due to the activity, crystalline states can undergo a drift instability and start to travel while keeping their spatial structure. Based on linear stability analyses, time simulations and numerical continuation of the fully nonlinear states, we present a detailed analysis of the bifurcation structure of resting and traveling states. We explore, for instance, how the slanted homoclinic snaking of steady localized states found for the passive PFC model is modified by activity. The analysis is carried out for the model in two spatial dimensions. Morphological phase diagrams showing the regions of existence of various solution types are presented merging the results from all the analysis tools employed. We also study how activity influences the crystal structure with transitions from hexagons to rhombic and stripe patterns. This in-depth analysis of a simple PFC model for active crystals and swarm formation provides a clear general understanding of the observed multistability and associated hysteresis effects, and identifies thresholds for qualitative changes in behavior.
\end{abstract}

\maketitle
 
\section{Introduction}
Pattern formation is a fascinating phenomenon observed in both nature and laboratory experiments and studied theoretically in a wide variety of fields~\cite{ball1999pattern, lam1998, pismen2006patterns}. 

In the case of macroscopic physical systems, one can usually distinguish between passive systems that are typically closed and develop towards thermodynamic equilibrium, and active or non-equilibrium systems that are open and develop under permanent energy flow. In the former, the resulting states may exhibit spatial patterns, e.g., crystalline structures, that can be related to self-assembly as typical structure lengths result directly from the properties of individual constituents. In contrast, in active systems the structures that occur are self-organized and dissipative. In this case typical structure lengths result from transport coefficients \cite{CrossHohenberg}. 

One prominent example of an active system is a system consisting of active particles or agents like bacteria, animals or artificial micro-swimmers \cite{wadaPRL99,rappelPRL83,vicsekPRE74,sumpter}. These agents are able to transform different forms of energy into self-propelled directed motion \cite{Marchetti.RevModPhys.85,BDLR2016rmp} and use various energy sources to drive an internal motor mechanism; hence they represent a non-equilibrium system driven by a continuous energy flow. Artificial micro-swimmers, for instance, turn chemical energy \cite{howsePRL99} or radiation like light \cite{PalacciScience,jiangPRL105} or ultrasound \cite{WangACSNano2012} into actively driven, self-propelled motion. Also, vibrated granular media in confined geometries are employed as good model systems for certain aspects of active matter \cite{GranularExcitonsNature1996,ArTs2003pre,WHDL2013prl,NaMR2006jsme}.

In non-equilibrium systems with a large number of active particles, intriguing collective phenomena arise. In particular, short- and long-range interactions between individual particles can result in alignment mechanisms leading to directional ordering (so-called \textit{polar ordering}) and synchronized motion of the self-propelled particles \cite{uchidaPRL106,golestanianSoftMatter7}. The resulting collective modes of motion are often referred to as \textit{swarming} \cite{Marchetti.RevModPhys.85}. Animals often form swarms for better protection from predators. Further proposed functions include social interaction \cite{abrahams1985fish}, enhanced foraging \cite{partridge1983tuna, petroff2014thiovulum} and increased efficiency of motion as often observed for birds \cite{fish1995ducklings}. 

One of the most famous approaches to collective motion is the Vicsek model \cite{vicsek1995PRL}, where each individual particle adapts to the average direction of motion in some neighborhood, in the presence of noise. In general, depending on the specific interactions between particles, their density and the driving strength (called in the following the \textit{activity}) one observes different regimes of clustering, ordering and motion that one may, in analogy to equilibrium behavior, call gas, liquid, liquid-crystalline and crystalline states \cite{BDLR2016rmp,MaVC2018arpc}. Much recent attention has focused on an actively driven condensation phenomenon, a motility-induced phase separation between a gaseous and a liquid state that arises purely due to self-propulsion \cite{Ginot2015prx,SSWK2015prl,CaTa2015arcmp}.

However, for certain particle interactions and/or at quite high densities, active particles can also form crystalline ordered states, in particular, resting \cite{Thar2002,Thar2005} or traveling \cite{PalacciScience,theurkauff2012prl,LibchaberPRL2015,ginot2018aggregation} patches with nearly crystalline order \cite{ToTR2005ap}. Different bacteria can form crystalline structures. In particular, rotating cells of \textit{Thiovulum majus}, a very fast and smoothly swimming, large and nearly spherical bacterium equipped with flagella \cite{garrity2006}, are attracted to each other owing to flow fields created by their rotation, and bacterial crystals form. The connected cells can pull nutrient-rich water towards the swarm by collective motion of the flagella \cite{petroff2014thiovulum}. When buoyancy forces are included another mechanism for structure formation becomes available, resulting in the phenomenon of bioconvection \cite{childress}.

These ``active crystals'' \cite{MenzelLoewen,MenzelOhtaLoewenPhysRevE.89} (also called ``flying crystals'' \cite{ToTR2005ap} or ``living crystals'' \cite{PalacciScience,MSAC2013prl,BDLR2016rmp}) have properties that differ from known passive crystalline clusters \cite{speckPRL112,speckPRL110}. The activity due to self-propulsion can change the critical temperature and density at which crystallization sets in and may even be necessary for crystalline clusters to emerge. Besides, activity can induce organized translational and rotational motion \cite{theurkauff2012prl,ReBH2013pre,Ginot2015prx,ginot2018aggregation}. Patches of rotating cells of \textit{Thiovulum majus} can also rotate as a whole.

Many particle-based models are studied that show resting, traveling and rotating, active, crystalline and amorphous clusters \cite{EbEr2003pspi,REES2008epjt,MSAC2013prl,NKEG2014prl} as well as cluster-crystals \cite{Menz2013jpm,DeLH2017njp}.  For instance, a systematic study of the interplay of a short-range attraction and self-propulsion in Brownian dynamics simulations shows that clusters form at low activity (due to attraction) as well as at high activity (motility-induced) with a homogeneous active fluid phase in between \cite{ReBH2013pre}.

Besides discrete models like Vicsek's, there exist a number of continuum models for active matter \cite{ToTR2005ap,Marchetti.RevModPhys.85,Menz2015prspl,RKBH2018pre}. An important example is the Toner-Tu model of swarming \cite{ToTu1995prl,TonerTu}. It represents a generalization of the compressible Navier-Stokes equations of hydrodynamics to systems without Galilei invariance, i.e., with preferred velocities. Recently, a simple active phase-field-crystal (aPFC) model has been proposed that describes transitions between the liquid state on the one hand, and resting and traveling crystalline states on the other \cite{MenzelLoewen,MenzelOhtaLoewenPhysRevE.89}, combining elements of the Toner-Tu theory and the (passive) phase-field-crystal (PFC) model.

The PFC model is an intensively studied microscopic mean field model for the dynamics of crystallization processes on diffusive time scales \cite{EmmerichPFC}. It was introduced by Elder and coworkers \cite{ElderGrantPRL88} and applies to passive colloidal particles as well as to atomic systems \cite{TGTP2009prl,ERKS2012prl}. Mathematically, it corresponds to the conserved Swift-Hohenberg (cSH) equation \cite{TARG2013pre} in the form of a continuity equation. In contrast to the PFC model, the classical Swift-Hohenberg (SH) equation represents non-conserved dynamics \cite{EGUW2018springer}. The SH equation is a standard equation for studying pattern formation close to the onset of a monotonic short-wave instability in systems without a conservation law, e.g., a Turing instability in reaction-diffusion systems or the onset of convection in a B\'enard system \cite{CrossHohenberg}. The cSH equation was first derived as the equation governing the evolution of binary fluid convection between thermally insulating boundaries \cite{KnoblochPRA1989}. In the PFC context, recent derivations from classical Dynamical Density Functional Theory (DDFT) of colloidal crystallization can be found in Refs.~\cite{EmmerichPFC,ARTK2012pre, ArcherPFC}. In the course of the derivation, the one-particle density of DDFT is shifted and scaled to obtain the order parameter field of PFC. For brevity, in the following we refer to the resulting order parameter as a ``density''. 

Both SH and PFC models represent gradient dynamics on the same class of energy functionals \cite{EGUW2018springer}. However, in the active PFC model the coupling between density and polarization (quantified by an activity parameter coupling the two fields) breaks the gradient dynamics structure. Thus sustained motion becomes possible. In fact, nonvariational modifications of the standard nonconserved SH equation have also been studied and are known to exhibit traveling states, although with different onset behavior \cite{KoTl2007c,HoKn2011pre,BuDa2012sjads}.

Thus far, the active PFC model has been employed mainly to investigate the linear stability of the liquid state with respect to the development of resting and traveling crystalline patterns and in the study of domain-filling resting and traveling crystals by direct time simulations in different geometries \cite{MenzelLoewen,MenzelOhtaLoewenPhysRevE.89,ChGT2016el,PVWL2018pre}. A bifurcation study of a spatially one-dimensional system was first provided in \cite{OphausPRE18}. The main purpose of the present work is to investigate these states and the transitions between them in two spatial dimensions, focusing on both spatially extended and spatially localized states. Our aim is to present a detailed analysis of the underlying bifurcation structure that can serve as a reference for similar analyses of other models describing active crystals in the future. This allows one to develop a clearer understanding of the observed multistability between different states and associated hysteresis effects as well as identifying critical threshold states for the occurrence of qualitative changes in behavior.

In the present paper we focus in particular on the transition from resting to traveling states that will turn out to occur via drift-pitchfork bifurcations. Drift-pitchfork bifurcations are widely studied in the literature and occur in many systems \cite{FaDT1991jpi,GGGC1991pra}. This includes the onset of motion of self-aggregating membrane channels \cite{LeNH2006prl}, drifting liquid column arrays \cite{BrFL2001el}, chemically-driven running droplets \cite{JoBT2005epje} and traveling localized states in reaction-diffusion systems~\cite{SOBP_PRL97,PiPRL01,driftbif_gurevich}. The onset of motion of localized structures is studied, for instance, in Refs.~\cite{krischerPRL73, OBSP1998pre,baerPRE64,akhmedievPRE53} while Refs.~\cite{MenzelOhtaLoewenPhysRevE.89,OSIPOV1996,PVWL2018pre} focus on domain-filling patterns. 

In the PFC and aPFC models, spatially localized states correspond to finite crystalline patches (i.e., patches of periodic states) that coexist with a liquid background (i.e., a homogeneous state). A great variety of resting localized states has been analyzed in detail for the PFC model in Ref.~\cite{TARG2013pre} where detailed bifurcation diagrams are given in the case of one spatial dimension (1D) while the two (2D) and three (3D)-dimensional cases are investigated via direct numerical simulations. An example of a bifurcation diagram in 2D is given in \cite{EGUW2018springer}. 

In general, localized states are frequently observed in experiments and models in various areas of biology, chemistry and physics \cite{MathBio,BioPatterns,coulletPRL84,ChemWaves,BurkeKnoblochLSgenSHe,liehr2013dissipative,purwins2010dissipative}. Examples range from localized patches of vegetation patterns \cite{MERON2004}, local arrangements of free-surface spikes of magnetic fluids just below the onset of the Rosensweig instability \cite{richterPRL94} to localized spot patterns in nonlinear optical systems \cite{SchaepersPRL2000} and oscillating localized states (oscillons) in vertically vibrating layers of colloidal suspensions \cite{LiouPRL1999}.

In the context of solidification as described by PFC models, localized states are observed in and near the thermodynamic coexistence region of liquid and crystalline states. Crystalline patches of various sizes and symmetry can coexist with a liquid environment depending on control parameters such as the mean density and undercooling \cite{RATK2012pre,TARG2013pre, EGUW2018springer}. For instance, as the mean density increases, the possible crystal patches increase in size as new density peaks (or ``bumps'', or ``spots'') are added at their boundary. Ultimately, the whole available domain is filled and the branches of localized states terminate on a branch of space-filling periodic states. Within their existence region, the localized states are organized within a ``snakes-and-ladders'' structure in the bifurcation diagram \cite{BurkeKnoblochSnakingChaos2007,SandstedeSnakes}. In conserved systems like the PFC model on a finite periodic domain this structure is slanted \cite{TARG2013pre, BoCR2008pre,Dawe2008sjads,LoBK2011jfm,PACF2017prf} but in nonconserved systems like the SH model it is aligned in the vertical \cite{K_IMA16,BurkeKnoblochSnakingChaos2007,ALBK2010sjads,LSAC2008sjads}. On nonperiodic domains the boundary conditions may substantially modify this behavior \cite{HoughtonKPRE2009,MercaderBAKPRE2009,kozyreffPRL2009}.

We use the active PFC model of Ref.~\cite{MenzelLoewen} to explore how the slanted snaking of steady localized states -- a characteristic feature of pattern-forming systems with a conserved quantity -- is modified by activity. This includes the question when and how resting localized states start to travel and whether and how they are destroyed by activity. We also explore whether traveling localized states can exhibit generic slanted snaking.

The paper is organized as follows. In Sec.~\ref{sec:gov_eq} we present the model equations, describe some of their elementary properties and outline the numerical approaches used to solve them. In Sec.~\ref{sec:LS} we study properties of spatially localized structures described by the model in both passive and actives cases, focusing on the transition to drift in the latter case. We also construct regime diagrams summarizing the parameter regions where different states are present. Section~\ref{crystals} focuses on related results for spatially extended states and the paper concludes with a brief discussion in Sec.~\ref{sec:summary} outlining future work.

\section{\label{sec:gov_eq}The model: governing equations}
The local state variables of the aPFC model as introduced in Ref.~\cite{MenzelLoewen} are the scalar order parameter field $\psi(\mathbf{r},t)$ (referred to in the following as a ``density'') and the vector order parameter field $\mathbf{\ensuremath{P}}(\mathbf{r},t)$ (referred to in the following as a ``polarization'') that describes the local ordering and direction of the active drive. Here $\mathbf{r}\in \Omega \subset \mathbb{R}^\mathrm{n}$, where $\Omega$ denotes the domain. The field $\psi(\mathbf{r},t)$ is conserved, i.e., $\int_{\Omega} \mathrm{d\mathbf{r}} \,\psi=0$ is constant in time, and specifies the modulation about a mean density $\bar{\psi}$ that itself encodes the deviation from the critical point \cite{EmmerichPFC}. The field $\mathbf{\ensuremath{P}}(\mathbf{r},t)$ is in general nonconserved.

The uncoupled dynamics of $\psi(\mathbf{r},t)$ and $\mathbf{\ensuremath{P}}(\mathbf{r},t)$ correspond to a purely conserved and a mixed non-conserved and conserved gradient dynamics on an underlying free energy functional $\mathcal{F}[\psi,\mathbf{P}]$, respectively. The functional contains no terms mixing the two fields and the coupling is purely nonvariational, i.e., no part of it can be written as gradient dynamics. The coupling maintains the conserved character of the $\psi$-dynamics, i.e., the evolution of $\psi$ follows a continuity equation $\partial_t\psi=-\nabla\cdot\mathbf{j}$, where $\mathbf{j}$ is a flux. The nondimensional evolution equations are \cite{MenzelLoewen}
\begin{align}
\partial_{t}\psi &=  \nabla^{2}\frac{\delta\mathcal{F}}{\delta\psi}-v_{0}\nabla\cdot\mathbf{P},\label{eq:gov1}\\
\partial_{t}\mathbf{P} &= \nabla^{2}\frac{\delta\mathcal{F}}{\delta\mathbf{P}}-D_{\mathrm{r}}\frac{\delta\mathcal{F}}{\delta\mathbf{P}}-v_{0}\nabla\psi,
\label{eq:gov2}
\end{align}
where $v_0$ is the coupling strength, also called an activity parameter or strength of self-propulsion. Physically speaking, $\mathbf{\ensuremath{P}}$ is subject to translational and rotational diffusion with $D_{\mathrm{r}}$ being the rotational diffusion coefficient. The functional $\mathcal{F}[\psi,\mathbf{P}]$ is the sum of the standard phase-field-crystal functional $\mathcal{F}_\mathrm{PFC}[\psi]$ \cite{ElderGrantPRL88, ElderGrantPRE70, EmmerichPFC} and an orientational part $\mathcal{F}_{\mathbf{P}}[\mathbf{P}]$,
\begin{equation}
\mathcal{F}=\mathcal{F}_\mathrm{PFC}+\mathcal{F}_{\mathbf{P}},
\label{eq:functional}
\end{equation}
with
\begin{equation}
 \mathcal{F}_\mathrm{PFC}[\psi] = \int \mathrm{d}\mathbf{r}\left\{ \frac{1}{2}\psi\left[\epsilon+\left(1+\nabla^{2}\right)^{2}\right]\psi+\frac{1}{4}(\psi+\bar{\psi})^{4}\right\}
 \label{eq:PFCfunctional}
\end{equation}
and 
\begin{equation}
\mathcal{F}_{\mathbf{P}}[\mathbf{P}]=\int \mathrm{d}\mathbf{r} \left(\frac{1}{2}C_1\mathbf{P}^{2}+\frac{1}{4}C_4\mathbf{P}^{4}\right).
\label{eq:functionalP}
\end{equation}
The functional (\ref{eq:functional}) encodes the phase transition between the liquid and crystal states \cite{EmmerichPFC,TFEKA2019}. It contains a negative interfacial energy density ( $\sim|\nabla\psi|^2$)  that favors the creation of interfaces, a bulk energy density and a stabilizing stiffness term ($\sim(\Delta\psi)^2$) -- this can be seen by partial integration. The parameter $\epsilon$ encodes temperature such that negative values of $\epsilon$ correspond to an undercooling of the liquid phase and result in crystalline (periodic) states for suitable mean densities $\bar{\psi}$, whereas positive values of $\epsilon$ result in a liquid (homogeneous) phase. The functional (\ref{eq:functionalP}) with $C_1<0$ and $C_2>0$ allows for spontaneous polarization (pitchfork bifurcation at  $C_1=0$). However, in our work we avoid spontaneous polarization and use $C_1>0$ with $C_2=0$ as also done in most of the analyses of Refs.~\cite{MenzelLoewen,MenzelOhtaLoewenPhysRevE.89,ChGT2016el}. With $C_1>0$ diffusion tends to reduce polarization.

Computing the variational derivatives of the energies (\ref{eq:PFCfunctional}) and (\ref{eq:functionalP}) and introducing the result in the governing equations~(\ref{eq:gov1})-(\ref{eq:gov2}) leads to the dynamical equations
\begin{align}
\partial_{t}\psi &= \nabla^{2}\left\{\left[\epsilon+\left(1+\nabla^{2}\right)^{2}\right]\psi+\left(\bar{\psi}+\psi\right)^{3}\right\}-v_{0}\nabla\mathbf{\cdot P}, \label{eq:dtpsi} \\
\partial_{t}\mathbf{P} &= C_1\nabla^{2}\mathbf{P} - D_{\mathrm{r}}C_1\mathbf{P}-v_{0}\nabla\psi. 
\label{eq:dtP}
\end{align}
By construction, Eq.~(\ref{eq:dtpsi}) preserves $\int_{\Omega}\mathrm{d}\mathbf{r}\,\psi\equiv 0$ while the assumption $C_2=0$ implies that Eq.~(\ref{eq:dtP}) preserves $\int_{\Omega}\mathrm{d}\mathbf{r}\,\mathbf{P}\equiv 0$. Moreover, the equations are nonvariational whenever $v_0\neq0$ and are invariant under the reflection
\begin{equation}
\kappa: (\mathbf{r}, \psi, \mathbf{P})\to (-\mathbf{r}, \psi, -\mathbf{P}).
\end{equation}
This symmetry permits the presence of steady, nondrifting solutions that are not left-right symmetric, provided they are $\kappa$-symmetric. To see this, suppose we seek a solution that is stationary in a frame moving with speed $c$ in the $x$-direction, i.e., ${\bf c}=c{\hat{\bf x}}$. In the moving frame we have
\begin{align}
0 =& \nabla^2\left\{\left[\epsilon+\left(1+\nabla^2\right)^{2}\right]\psi+\left(\bar{\psi}+\psi\right)^{3}\right\}\nonumber \\ 
-&v_{0}\nabla\cdot\mathbf{P}  + \mathbf{c} \cdot \nabla \psi, \label{eq:steadystatePSI2d}\\ 
\mathbf{0} =& C_1 \nabla^{2}\mathbf{P}-D_{\mathrm{r}}C_1 \mathbf{P} - v_{0}\nabla\psi + (\mathbf{c}\cdot\nabla) \mathbf{P}. \label{eq:steadystateP2d}
\end{align}
Suppose now that the solution $(\psi,\mathbf{P})$ is $\kappa$-symmetric with respect to $x\to -x$. Applying $\kappa$ to (\ref{eq:steadystatePSI2d})-(\ref{eq:steadystateP2d}) we obtain
\begin{align}
0 =& \nabla^2\left\{\left[\epsilon+\left(1+\nabla^2\right)^{2}\right]\psi+\left(\bar{\psi}+\psi\right)^{3}\right\} \nonumber \\ 
-&v_{0}\nabla\cdot\mathbf{P} - \mathbf{c} \cdot \nabla \psi,\\
\mathbf{0} =& C_1 \nabla^{2}\mathbf{P}-D_{\mathrm{r}}C_1 \mathbf{P} - v_{0}\nabla\psi - (\mathbf{c}\cdot\nabla) \mathbf{P}.
\end{align}

Together these equations imply that ${\bf c}\equiv 0$ and hence that a $\kappa$-symmetric solution is necessarily at rest. In the following we refer to such solutions as {\it resting} solutions. Note that $\kappa$ symmetry is a robust condition for a resting state. However, Eqs. (\ref{eq:dtpsi})--(\ref{eq:dtP}) also admit robust resting states that are not $\kappa$-symmetric (see below). Such states are present here because of the special structure of the equations and would not be present in generic models except at isolated parameter values. Each of these states may in turn undergo transitions to a drifting state as parameters are varied, and in the remainder of this paper we focus on the properties of both resting and traveling states in 1D and 2D with a special emphasis on the onset of motion that arises from spontaneous breaking of the $\kappa$ symmetry.

Resting and traveling solutions of the aPFC model (\ref{eq:dtpsi})--(\ref{eq:dtP}) in 1D were studied in detail in Refs.~\cite{OphausPRE18,OKGT2020arxiv}. However, in nature, collective motion often occurs effectively in 2D. Tissue cells, bacteria and amoebae crawl on substrates while ungulates like gnu or sheep display herding and organize in 2D swarms. In the context of artificial active matter, e.g., colloidal particles swimming on the surface of a liquid, the system may form 2D crystals \cite{PalacciScience}. For this reason, we investigate here how 2D active crystals described by the aPFC model evolve from a localized state (LS) consisting of a single peak into spatially extended states (crystals) under the influence of activity. As in \cite{OphausPRE18}, we focus on the mean density $\bar{\psi}$ and the activity parameter $v_0$ as the main control parameters. The activity parameter $v_0$ must, of course, be nonzero for the presence of traveling structures but its specific value will turn out to have a major influence not only on the transition from resting to traveling states but also on the structure of 2D crystals and associated pattern selection.

\subsection{\label{sec:num2d}Numerical continuation in 2D}

\begin{figure}
\centering\hspace*{-0.cm}
\includegraphics[width=0.5\textwidth]{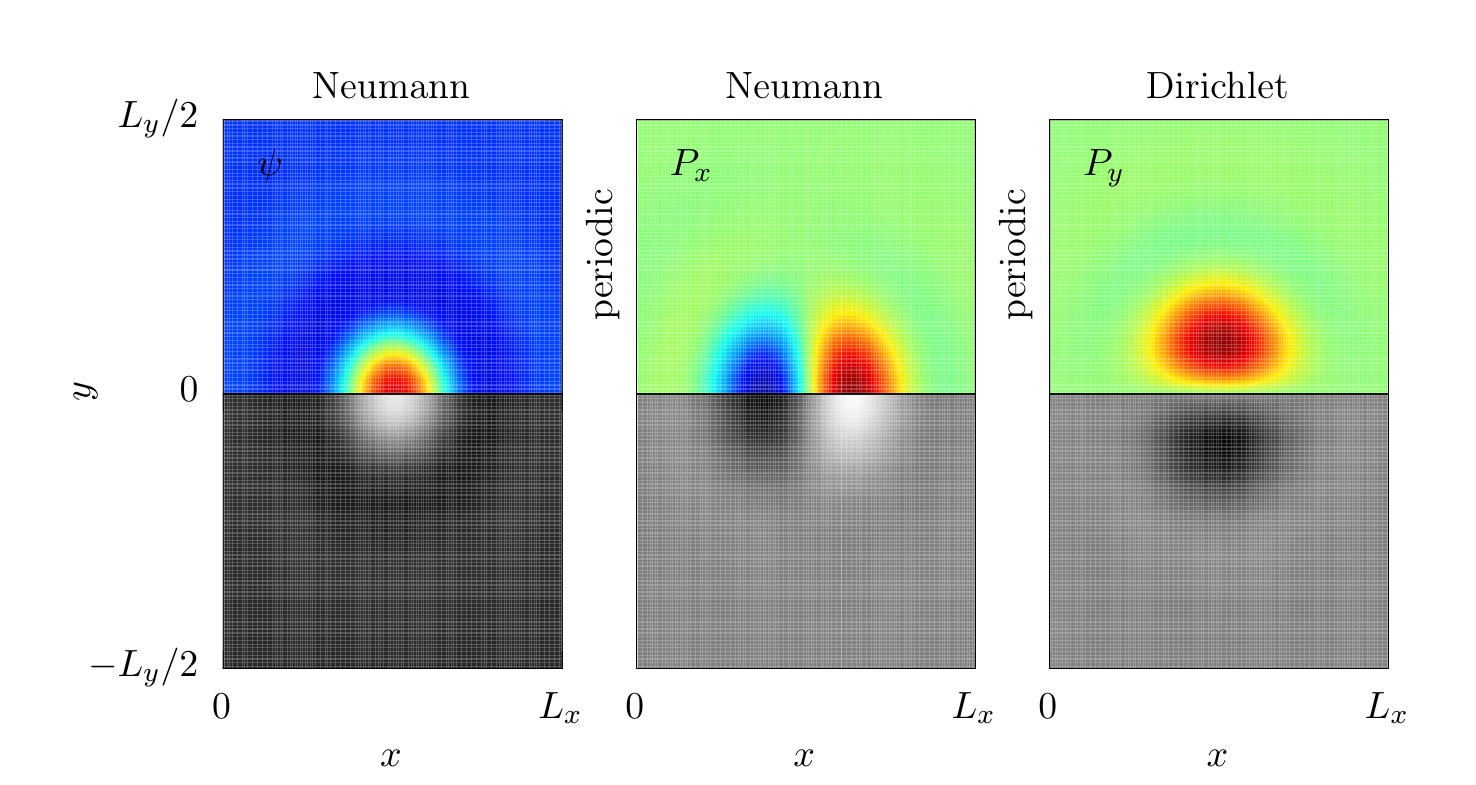}	
 \caption{Sketch of the numerical setup: Two-dimensional structures that are reflection-symmetric with respect to the $x$-axis are computed on a reduced domain $[0,L_x] \times [0,L_y/2]$ as indicated by the colored area. The density field $\psi$ and the $x$-component $P_x$ of the polarization satisfy Neumann boundary conditions at $y=0$ while Dirichlet boundary conditions apply to the $y$-component $P_y$ of the polarization: $P_y=0$ at $y=0$. All fields are periodic in the $x$-direction, i.e., only motion in the $x$-direction is allowed. For visualization, the entire domain $\Omega_\mathrm{exp}= [0,L_x] \times [-L_y/2,L_y/2]$, indicated by the colored and gray-scale regions, is used.}
 \label{fig:BC}
\end{figure}

We employ numerical parameter continuation \cite{KrauskopfOsingaGalan-Vioque2007,DWCD2014ccp,EGUW2018springer} to determine steady ($c=0$) and stationary ($c>0$) periodic and localized solutions of Eqs.~(\ref{eq:steadystatePSI2d}) and (\ref{eq:steadystateP2d}). We use the MATLAB package PDE2PATH \cite{uecker_wetzel_rademacher_2014} which allows us to follow branches of solutions in parameter space, detect bifurcations, switch branches and in turn follow the bifurcating branches. A phase condition that breaks translational invariance and a constraint that enforces the mean density $\bar{\psi}$ are included as integral conditions. This implies that in each continuation run beside the main control parameter one has two auxiliary parameters that have to be adapted. The mean density $\bar{\psi}$ and the activity $v_{0}$ are used as the main control parameters while the velocity $c$ and a Lagrange multiplier for the density constraint are the auxiliary parameters that are adapted. All solutions satisfy in addition the condition $\int_{\Omega}\mathrm{d}\mathbf{r}\,\mathbf{P}\equiv 0$.

Since 2D computations are much more expensive and time-consuming as compared to 1D problems we make use of the symmetries of the fields $\psi$ and $\mathbf{P}=(P_x,P_y)^T$ to reduce the computational effort. Unless otherwise stated in the caption of the figures that follow, all computations are carried out on the half-domain as explained in Fig.~\ref{fig:BC}. Here the colored area $\Omega=[0,L_x] \times [0,L_y/2]$ indicates the part of the domain on which the actual computation is performed. The entire solution profile is then obtained by exploiting the following symmetries:
\begin{align}
\psi(x,y)=&\psi(x,-y),\\
P_x(x,y)=&P_x(x,-y),\quad P_y(x,y)=-P_y(x,-y), 
\nonumber                                                                                                                                                                                                                                            \end{align}
where $y=0$ corresponds to the horizontal line separating the colored and gray areas. The expanded area used for visualization and classification of the solutions is thus $\Omega_\mathrm{exp}=[0,L_x] \times [-L_y/2, L_y/2]$ with volume or area $V=L_x\,L_y$. From the linear stability analysis of the uniform state, we know that at onset only the unstable mode $k_\mathrm{c}=1 \Leftrightarrow L_\mathrm{c}=2\pi$ grows.

Next, we define the boundary conditions (BCs) imposed on $\Omega=[0,L_x] \times [0,L_y/2]$. In order to pin the solutions such that the applied symmetries are preserved, we use Neumann BCs in the $y$-direction for $\psi$ and $P_x$. Accordingly, $P_y$ is kept zero at $y=0$ and $y=L_y/2$, i.e., Dirichlet BCs are applied. The combined BCs in the $y$-direction read
\begin{align}
\partial_y\psi(x,y=0,L_y/2)=&0,\\
\partial_y P_x(x,y=0,L_y/2)=&0,\quad P_y(x,y=0,L_y/2)=0.\nonumber                                                                                                                                                                                                                                            \end{align}
In the $x$-direction, periodic BCs are imposed on all three fields.

Owing to the chosen BCs, the $y$-component $c_y$ of the drift velocity $\mathbf{c}$ always remains zero. This implies that crystalline structures have to be oriented such that the desired drift, e.g., as observed in time simulations or experiments, is in the $x$-direction, i.e., $c_x\neq0$.   

Besides the rectangular geometry, we also make use of a hexagonal domain when discussing the passive PFC model and the phenomenon of slanted snaking of branches of steady LS. There, the numerical continuation is done on a triangular domain, namely, a right-angled triangle with a hypotenuse of the side length of the hexagon and Neumann BCs for $\psi$. In the passive case, $P_x$ and $P_y$ remain zero. The triangle defined by the vertices at $(x,y) = 2\pi(0, 0),\,2\pi(0, 3)$ and $2\pi(1, 3/\sqrt{3})$ is one twelfth of the entire domain as pictured in Fig.~\ref{fig:hex_snaking_eps098}. Note that the equilateral triangles found in the hexagon have a height $L_\mathrm{c}=2\pi$ and a side length $L_\mathrm{a}=\frac{2}{\sqrt{3}}L_\mathrm{c}=\frac{4\pi}{\sqrt{3}}$.

All the bifurcation diagrams that follow show the $L^2$-norm of the density profile that we use as the main solution measure. In 2D, this norm is defined by
\begin{align}
 ||\psi||_2=\sqrt{\frac{1}{V}\int_V \mathrm{d}\mathbf{r}\,\psi(\mathbf{r})^2}
\end{align}
with area $V$ and $\mathbf{r}=(x,y)^T \in V \subset \mathbb{R}^2$.
In addition to numerical continuation, we also perform numerical time simulations employing a pseudo-spectral method with semi-implicit Euler time-stepping.

\section{\label{sec:LS}Localized states}

As known from the passive PFC model \cite{TARG2013pre} and from results in 1D \cite{OphausPRE18}, we can identify a transition region where patches of the liquid and crystalline states coexist. In the vicinity of the linear instability threshold of the liquid state, a broad variety of spatially localized states (LS or crystallites) is therefore expected.  

We use numerical continuation of Eqs.~(\ref{eq:steadystatePSI2d}) and (\ref{eq:steadystateP2d}) to explore the bifurcation structure of the resulting active crystallites in 2D. How do (active) crystallites grow in the plane as a function of the mean density $\bar{\psi}$? What is the influence of the activity parameter $v_0$? Are fully 2D traveling states possible? Do traveling LS exhibit the same slanted snaking as the resting LS?

\subsection{\label{sec:passive_snaking}Passive PFC model: slanted snaking}
	
	\begin{figure}
	\centering\hspace*{-0.3cm} 
     \includegraphics[width=0.5\textwidth]{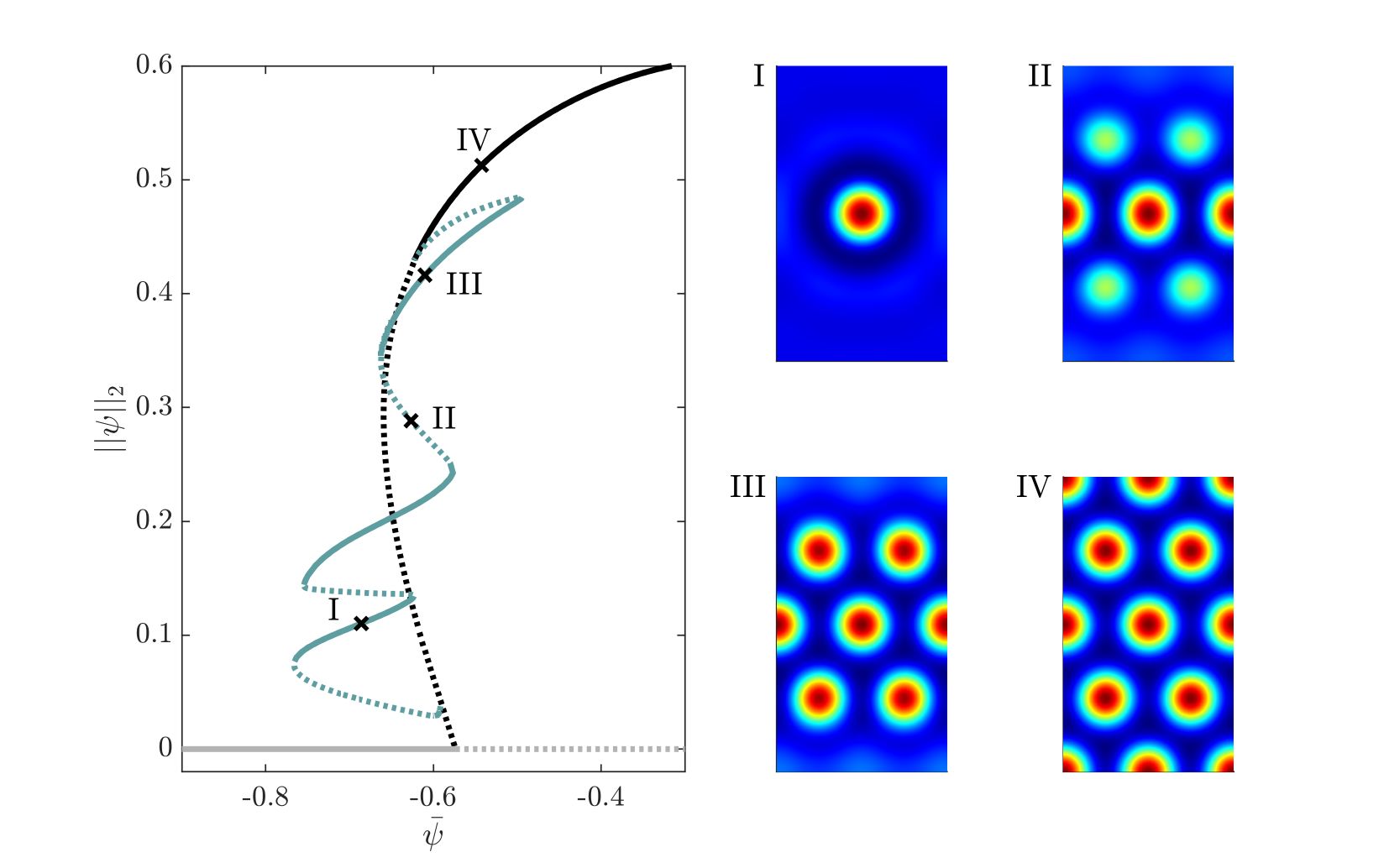}	
 	 \caption{(left) Bifurcation diagram showing branches of homogeneous, periodic and localized steady states of the passive PFC model ($v_0=0$) on a rectangular domain. Shown is the $L^2$-norm $||\psi||_2$ as a function of the mean density $\bar{\psi}$.  Stable and unstable states are shown as solid and dotted lines, respectively. The liquid phase (gray horizontal line) is destabilized at $\bar{\psi}\approx-0.55$ and an unstable branch of periodic hexagonal patterns (black line, cf.~location IV) emerges subcritically. In a secondary bifurcation, a branch of LS (blue line) is created. After various folds responsible for repeated gain and loss of stability, the LS branch terminates on the same branch of periodic hexagons from which it bifurcated. (right) Selected solution profiles $\psi(\mathbf{r})$ at locations labeled I to IV in the left panel. The domain size is $2L_\mathrm{a} \times 4L_\mathrm{c}$ with $L_\mathrm{a}=2L_\mathrm{c}/\sqrt{3}$ being the side length of a hexagon/triangle and $L_\mathrm{c}=2\pi$ the critical wavelength/height of the triangles. The remaining parameter is $\epsilon=-0.98$.}
	 \label{fig:rect_snaking_eps098}
	 \end{figure}
	
	\begin{figure}
	\centering\hspace*{-0.3cm}
     \includegraphics[width=0.5\textwidth]{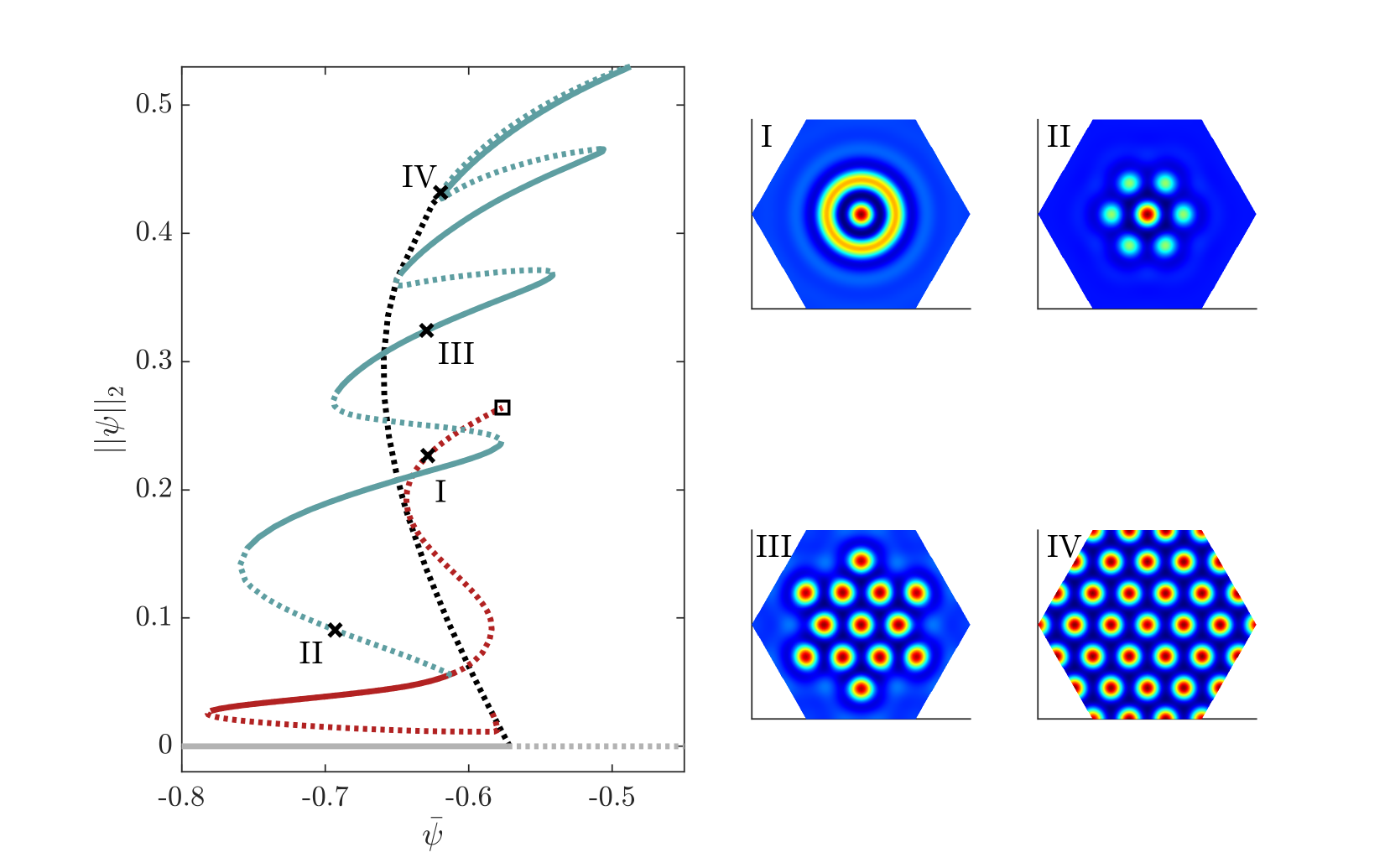}
	 \caption{(left) Bifurcation diagram showing branches of homogeneous (gray line), periodic (black line), target-like (red line) and localized (blue line) steady states of the passive PFC model ($v_0=0$) on a hexagonal domain. Shown is the norm $||\psi||_2$ as a function of the mean density $\bar{\psi}$. 
           (right) Selected solution profiles $\psi(\mathbf{r})$ at locations labeled I to IV in the left panel. The hexagonal domain has side length $3L_\mathrm{a}$. Remaining line styles and parameters are as in Fig.~\ref{fig:rect_snaking_eps098}.}
	 \label{fig:hex_snaking_eps098}
	 \end{figure}

	 \begin{figure}[t]
	\centering\hspace*{-0.0cm}
	 \includegraphics[width=0.5\textwidth]{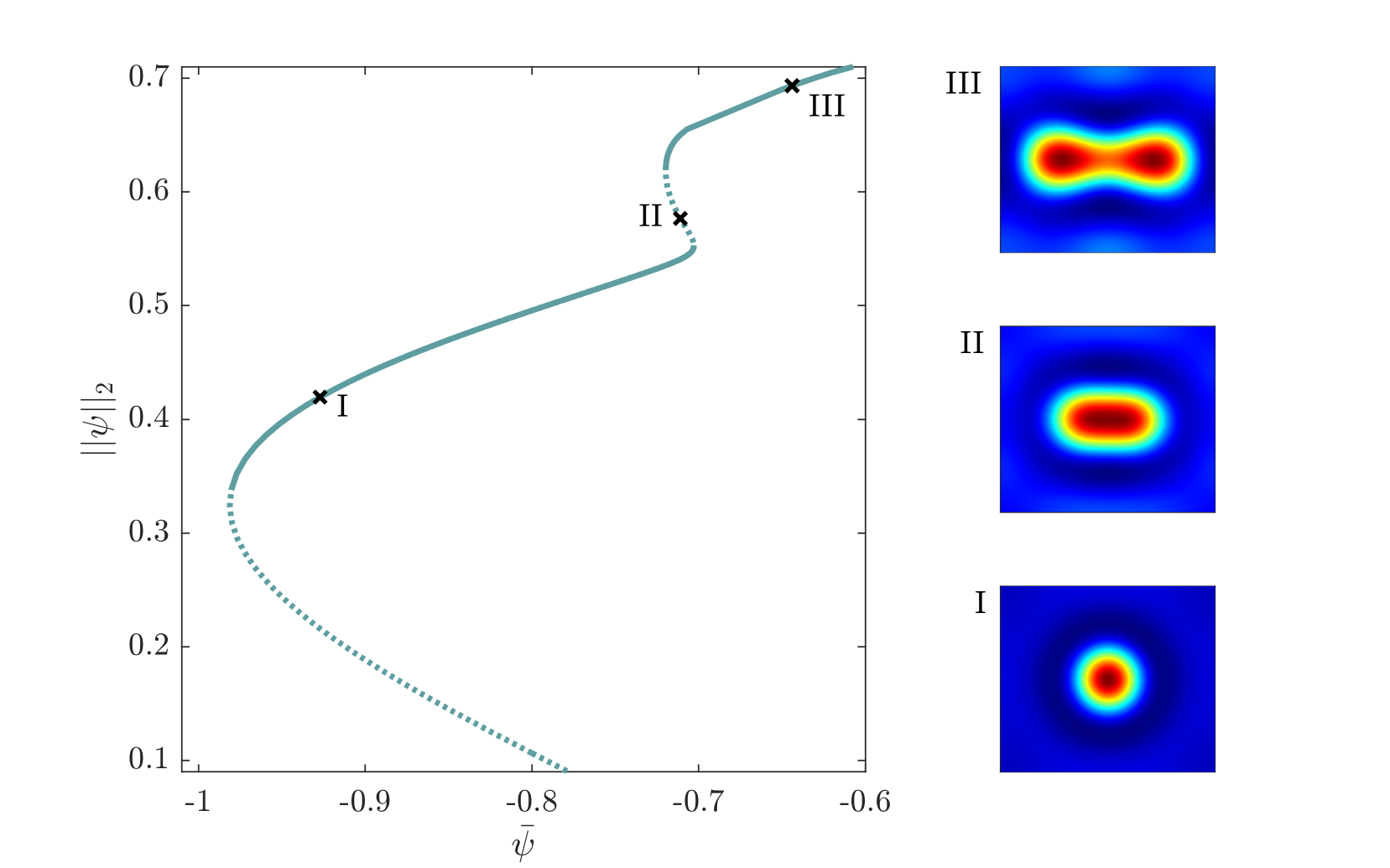}
 	 \caption{(left) Bifurcation diagram showing suppressed snaking at lower values of $\epsilon$, here $\epsilon=-1.5$, in the passive PFC model ($v_0=0$). In contrast to $\epsilon=-0.98$ (Figs.~\ref{fig:rect_snaking_eps098} and \ref{fig:hex_snaking_eps098}) the crystalline patches do not grow by adding layers of density peaks. Instead a single peak grows into an elongated structure and subsequently forms a dumbbell-shaped two-peak state. Asymmetric states are omitted.
         The domain size is $2L_\mathrm{a} \times 2L_\mathrm{c}$.
         Remaining line styles and parameters are as in Fig.~\ref{fig:rect_snaking_eps098}.}
	 \label{fig:no_snaking_eps15}
	 \end{figure}

We start by constructing bifurcation diagrams as a function of the mean density $\bar{\psi}$ for the passive PFC model, i.e., by setting $v_0=0$, resulting in uncoupled Eqs.~(\ref{eq:steadystatePSI2d}) and (\ref{eq:steadystateP2d}), with $\mathbf{P}\equiv\mathbf{0}$ for all time. 

Figures~\ref{fig:rect_snaking_eps098} and \ref{fig:hex_snaking_eps098} depict typical slanted snaking of the LS branches along their path to a spatially extended crystal. In both bifurcation diagrams, the continuation in $\bar{\psi}$ starts from the uniform state $\psi=0$ referred to as the \textit{liquid} state (gray branch). At a critical density close to $\bar{\psi}=-0.55$, this state loses stability and a branch of periodic solutions of hexagonal order bifurcates (black branch) in a transcritical bifurcation. We did not follow the supercritical part of the emerging branch that corresponds to so-called cold or down-hexagons. 

As expected, a secondary bifurcation is detected on the branch of periodic states close to the primary bifurcation. On the rectangular domain used in Fig.~\ref{fig:rect_snaking_eps098}, the bifurcating branch (blue line) corresponds to spatially localized hexagonal crystallites. The branch undergoes a series of folds corresponding to the addition of a pair of layers of density peaks, symmetrically with respect to $y=0$ (Fig.~\ref{fig:rect_snaking_eps098}); this is not the case in larger domains, however \cite{TFEKA2019}. At each fold, the stability of the branch changes. Solid lines correspond to stable solutions and dotted lines indicate unstable solutions. Eventually, the LS branch terminates on the branch of the spatially extended hexagons and the entire domain is filled with the crystalline state. Thereafter, the crystalline state is stable. Owing to the conservation of $\psi$, the loci of the left and right saddle-node bifurcations align along lines slanted towards higher $\bar{\psi}$. Since the model is passive with $v_0=0$ and $\mathbf{P}\equiv\mathbf{0}$, no traveling states can exist and all solutions are steady. 

Figure~\ref{fig:hex_snaking_eps098} presents a similar bifurcation diagram obtained from continuation on a hexagonal domain. In contrast to the rectangle used in Fig.~\ref{fig:rect_snaking_eps098}, a rotationally symmetric solution (red line, location I) emerges at the first secondary bifurcation from the hexagonal state. This type of LS has been termed a \textit{ring} solution. Its branch has been tracked until the state starts to interact with the Neumann boundaries of the triangular computation domain and its symmetry is destroyed. Apparently, the hexagonal geometry favors the emergence of ring-like solutions as it is closer to rotational symmetry than the previously used rectangle. 

The snaking branch of LS (blue line) bifurcates in a tertiary bifurcation from the branch of ring solutions. This bifurcation is actually imperfect due to numerical grid effects. However, in Fig.~\ref{fig:hex_snaking_eps098}, this cannot be seen by eye. The crystalline patch of hexagonal order gradually grows until the hexagonal domain is completely filled and the branch terminates on the branch of the periodic crystals (black line). As the hexagonal domain is of a larger area than the rectangular one used in Fig.~\ref{fig:rect_snaking_eps098}, more density peaks fit in and the LS snaking branch consists of more back-and-forth oscillations. Note, in particular, that in both cases the LS, be they hexagonal patches or rings, are present below the fold of the spatially extended hexagonal state, i.e., outside of the region of bistability between the trivial state and the hexagonal crystal. This observation confirms that the coexistence region is wider than the region of bistability -- a typical feature of systems with a conserved quantity.

Figures~\ref{fig:rect_snaking_eps098} and \ref{fig:hex_snaking_eps098} use $\epsilon=-0.98$ as employed in earlier studies \cite{MenzelLoewen, MenzelOhtaLoewenPhysRevE.89, PVWL2018pre}. However, at yet smaller values of the temperature-like parameter $\epsilon$, e.g., $\epsilon=-1.5$, the localized density peak does not grow into a patch of hexagonal order but rather elongates, forming first an oval structure and ultimately a dumbbell state (Fig.~\ref{fig:no_snaking_eps15}). On rectangular domains this elongation is a natural consequence of the domain shape and represents a continuous transition. However, this elongated state is not a consequence of boundaries: the continuation was carried out on various domains with the same result of an elongating density peak. In particular, and in contrast to all other solution profiles shown here, $\psi(\mathbf{r})$ in Fig.~\ref{fig:no_snaking_eps15} is not computed on half of the depicted domain and mirrored, and so states (I)-(III) depict the actual computed solution profiles. Here, the density peak is placed in the middle of the computational domain in order to avoid a possible influence of the boundaries. Based on these computations we conclude that the observed states describe gradual spot fission as $\bar{\psi}$ varies, i.e., fission of a spot into a pair of adjacent spots (see, e.g., \cite{kolokolnikov}). We have found no evidence for the coexistence of this state with any spatially extended state at these parameter values. Note that dumbbell localized states were previously observed in the classical nonconserved SH equation in both 2D and 3D~\cite{BC-PRE-05}.

Next, we move on to the active PFC model and investigate the influence of the activity parameter $v_0$. By continuation in $\bar{\psi}$ at $v_0=0$, we produce various LSs whose response to activity is then studied. As explained in Sec.~\ref{sec:num2d}, we use a rectangular domain and symmetries of $\psi$ and $\mathbf{P}$ in order to perform continuation on a reduced-size domain. In Sec.~\ref{sec:activesnaking}, we return to slanted snaking and study to what extent snaking is modified by activity. In particular, we study the bifurcation structure of traveling states as a function of $\bar{\psi}$.

\subsection{\label{sec:v0}Active PFC model: onset of motion}

\begin{figure}	
	\centering\hspace*{-0.3cm}
     \includegraphics[width=0.5\textwidth]{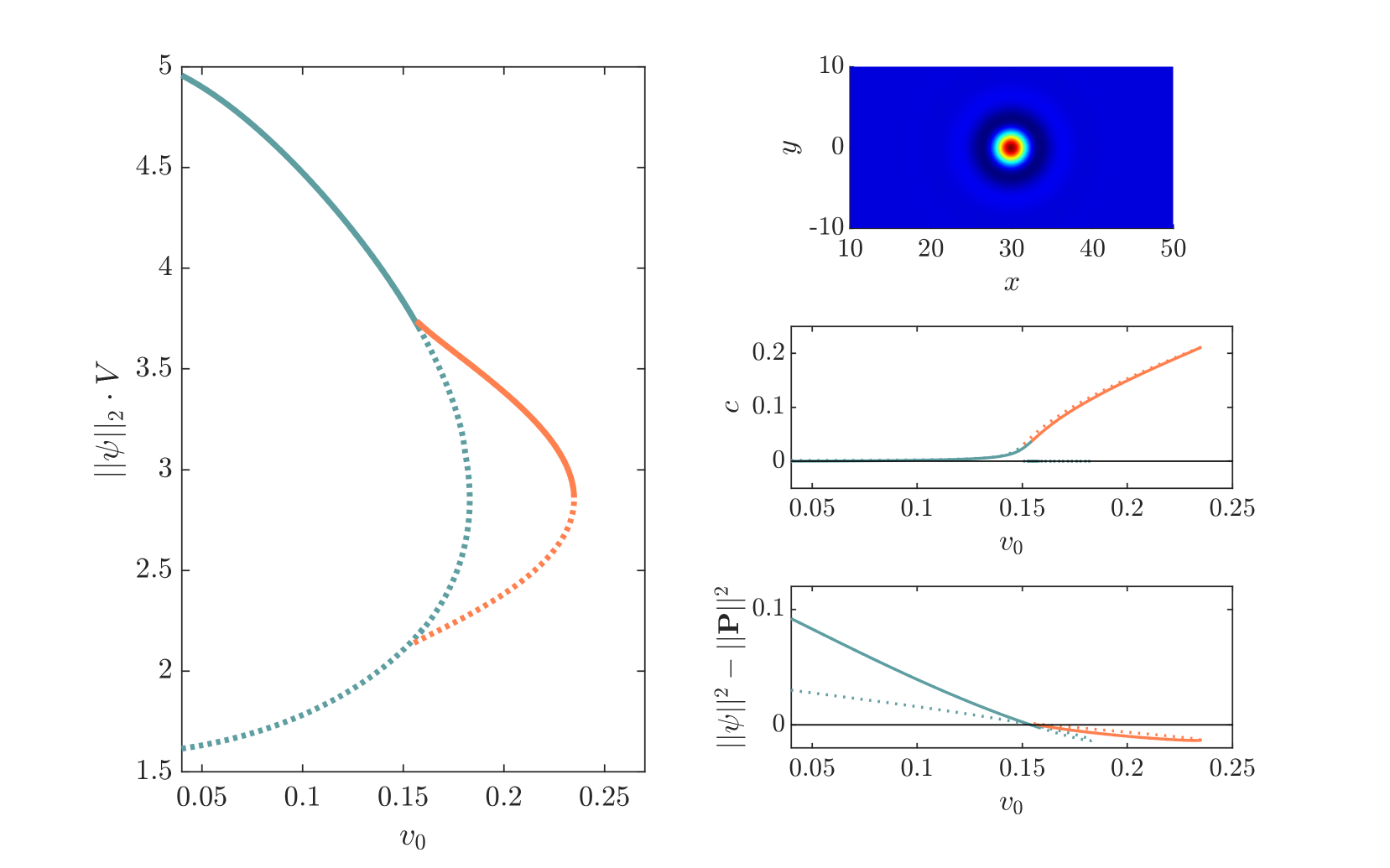}
	 \caption{(left) Bifurcation diagram of resting and traveling one-peak LS at mean density $\bar{\psi}=-0.9$ showing the $L^2$-norm of $\psi$ as a function of the activity $v_0$. Resting and traveling states are indicated by blue and orange lines, respectively. Branches of stable and unstable states are shown as solid and dotted lines. At $v_{\mathrm{c}}\approx0.15$, a stable resting LS undergoes a drift pitchfork bifurcation and a branch of traveling localized states (TLS) emerges. The region of existence of the TLS is limited by a fold at $v_0\approx0.24$. The panels on the right show (top) selected solution profile $\psi(\mathbf{r})$ at $v_0=0.1$ (only part of the domain is shown); (center) the drift velocity $c$ vs.\ $v_0$. Above $v_{\mathrm{c}}\approx0.15$, the velocity increases as $\sqrt{v_0-v_{\mathrm{c}}}$. Deviations from a sharp onset of motion are due to lattice effects. (bottom) The difference $||\psi||_2^2-||\mathbf{P}||_2^2$ crosses zero at the drift pitchfork bifurcation. Note that, in the left panel, $||\psi||_2$ times the area $V$ is plotted for clarity as for 2D domains the norm of LS tends to be very small. The domain size is $V=60\times30$. Remaining parameters are $\epsilon=-1.5$, $C_1=0.1$, $C_2=0$ and $D_\mathrm{r}=0.5$.
	  }
	 \label{fig:singlebump}
	 \end{figure}

	 \begin{figure}[t]
	 \centering
     \includegraphics[width=0.5\textwidth]{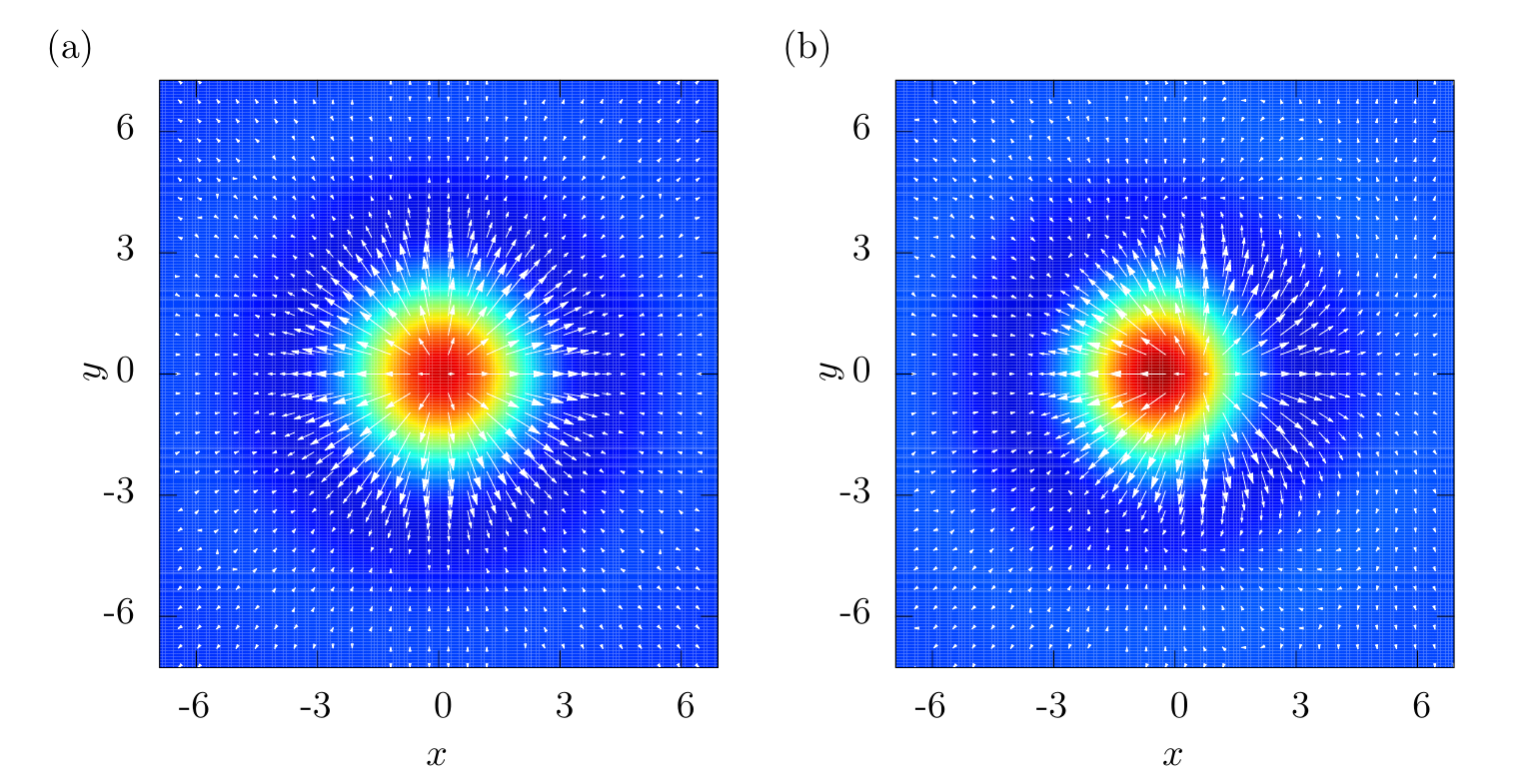}
	  \caption{Density $\psi(\mathbf{r})$ and polarization $\mathbf{P}(\mathbf{r})$ profile shown as a color map overlaid with white arrows for (a) a resting ($v_0=0.13$) and (b) a traveling $(v_0=0.22>v_\mathrm{c})$ one-peak LS. Note that in (a) the +1 defect of the polarization field coincides with the density maximum (and the net polarization is zero), while in (b) they are shifted with respect to one another as the front-back symmetry is broken. The shift corresponds to a net polarization, i.e., net propulsion to the left, with $c\approx-0.19$. Only a part of the computational domain is shown. The remaining parameters are as in Fig.~\ref{fig:singlebump}. 
	  }
	  \label{fig:pol_LS}
	 \end{figure}

	\begin{figure}[t]
	\centering
     \includegraphics[width=0.5\textwidth]{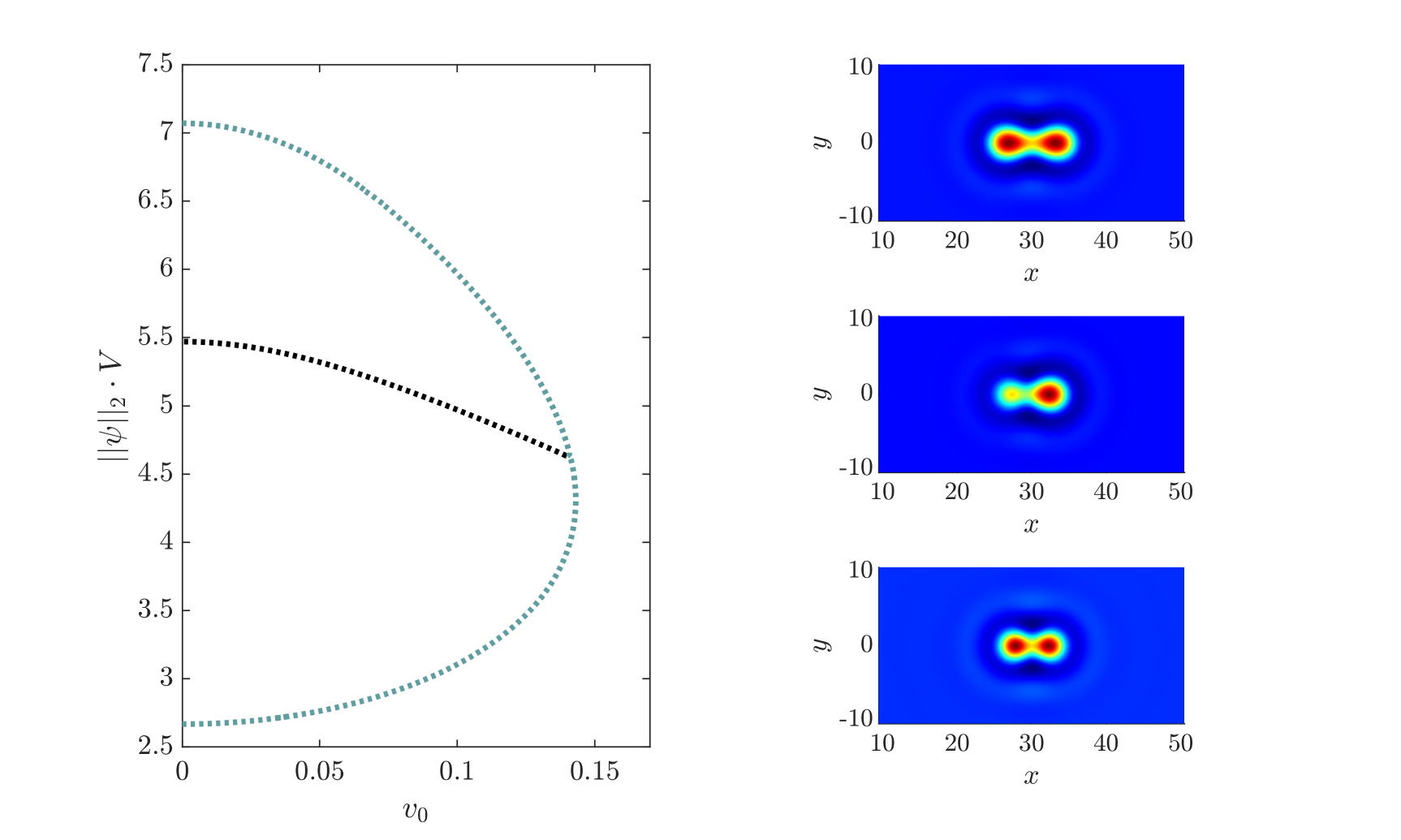}
	 \caption{ (left) Bifurcation diagram of a resting two-peak LS (cf.~Fig.~\ref{fig:no_snaking_eps15}) at mean density $\bar{\psi}=-0.9$ showing the $L^2$-norm of $\psi$ as a function of the activity $v_0$. (right) Selected density profiles $\psi(\mathbf{r})$ at $v_0=0.03$ (top) on the upper part of the branch of left-right symmetric states (blue line), $v_0=0.09$ (middle) on the branch of left-right asymmetric states (black line) and $v_0=0.04$ (bottom) on the lower part of the branch of the left-right symmetric states. In contrast to the one-peak LS, only resting two-peak LS exist at this mean density as the saddle-node bifurcation of the symmetric states is located at $v_0<v_{\mathrm{c}}\approx0.15$. The remaining line styles, parameters and the domain size are as in Fig.~\ref{fig:singlebump}.} 
	 \label{fig:restingdoublebump}
	\end{figure}

We now systematically explore how LS in 2D respond to increasing activity by employing the activity parameter $v_0$ as the main control parameter. From results obtained for LS in 1D~\cite{OphausPRE18}, we expect transitions from resting to traveling LS (RLS and TLS, respectively) associated with symmetry breaking between the two fields $\psi$ and $\mathbf{P}$, as centers of the density peaks shift with respect to +1 defects in $\mathbf{P}$ at a critical activity $v_\mathrm{c}$. For resting crystals, $\mathbf{P}$ points down the gradient of $\psi$, leading to a defect at the center of the density peak, similar to the vector field of a monopole. These defects are termed +1 defects [cf.~Fig.~\ref{fig:pol_LS}(a)].

Figure~\ref{fig:singlebump} shows a typical bifurcation diagram as a function of $v_0$. A stable one-peak LS at rest (represented by a solid blue line) undergoes a drift-instability at $v_\mathrm{c}\approx0.15$. The traveling one-peak LS (orange branch) are stable up to $v_0\approx0.24$ where the branch folds back to smaller $v_0$. Unstable branches are shown as dotted lines. The drift velocity $c$ of the TLS increases as $\sqrt{v_0-v_{\mathrm{c}}}$ as previously observed. Due to larger grid effects in 2D (we use an adaptive grid), the onset of motion is not perfectly sharp in $c$ (cf. Fig.~\ref{fig:singlebump}). Since the criterion for the onset of motion derived in~\cite{OphausPRE18} applies in two spatial dimensions, we track the quantity $||\psi||_2^2-||\mathbf{P}||_2^2$ to reveal a zero crossing at $v_\mathrm{c}$, and use this procedure to identify drift bifurcations. 

The onset of motion is associated with the appearance of symmetry breaking between $\psi$ and the vector field $\mathbf{P}$ for sufficiently large $v_0$. Centers of the density peaks shift with respect to the +1 defects in $\mathbf{P}$ as depicted in Fig.~\ref{fig:pol_LS}. For resting states, averaging $\mathbf{P}$ over the area of a single density peak yields zero. Above $v_\mathrm{c}$, a net orientation of $\mathbf{P}$ emerges and traveling crystals or crystallites come into existence. In Fig.~\ref{fig:pol_LS}(b), the net polarization points to the left leading to a negative drift velocity $c$. The direction of the shift and hence the resulting sign of the velocity are arbitrary: both directions correspond to the same branch of traveling solutions. This agrees well with similar observations for the onset of motion for extended patterns \cite{MenzelLoewen,PVWL2018pre}.

If $\bar{\psi}$ is chosen too low, i.e., too close to the solid-liquid transition, activity can melt crystallites before motion sets in. This is what happens to two-peak LS at rest at $\bar{\psi}=-0.9$ as $v_0$ increases and the two-peak LS passes through a fold before encountering a parity-breaking bifurcation (Fig.~\ref{fig:restingdoublebump}). Here the branch of two-peak LS does not reach far enough in $v_0$ to fulfill the criterion for the onset of motion and activity melts the structure before the onset of drift: the position of the fold is at $v_0\approx0.14<v_{\mathrm{c}}$. Close to the fold, there is a subcritical pitchfork bifurcation generating steady but asymmetric solutions [dotted black branch, cf.~Fig.~\ref{fig:restingdoublebump} (right central panel)] that bifurcate off the blue branch corresponding to solutions with left-right symmetry in $\psi(\mathbf{r})$ (right upper and lower panels). Note that the dotted black line represents two different asymmetric solutions related by reflection with respect to a suitable origin: $\left[\psi(x,y),\,P_x(x,y),\,P_y(x,y)\right] \leftrightarrow \left[\psi(-x,y),\,-P_x(-x,y),\,P_y(-x,y)\right]$. At $\bar{\psi}=-0.9$ these two-peak LS coexist with the one-peak LS from Fig.~\ref{fig:singlebump} but all two-peak states are unstable.

	\begin{figure}[t]
	\centering
	 \includegraphics[width=0.45\textwidth]{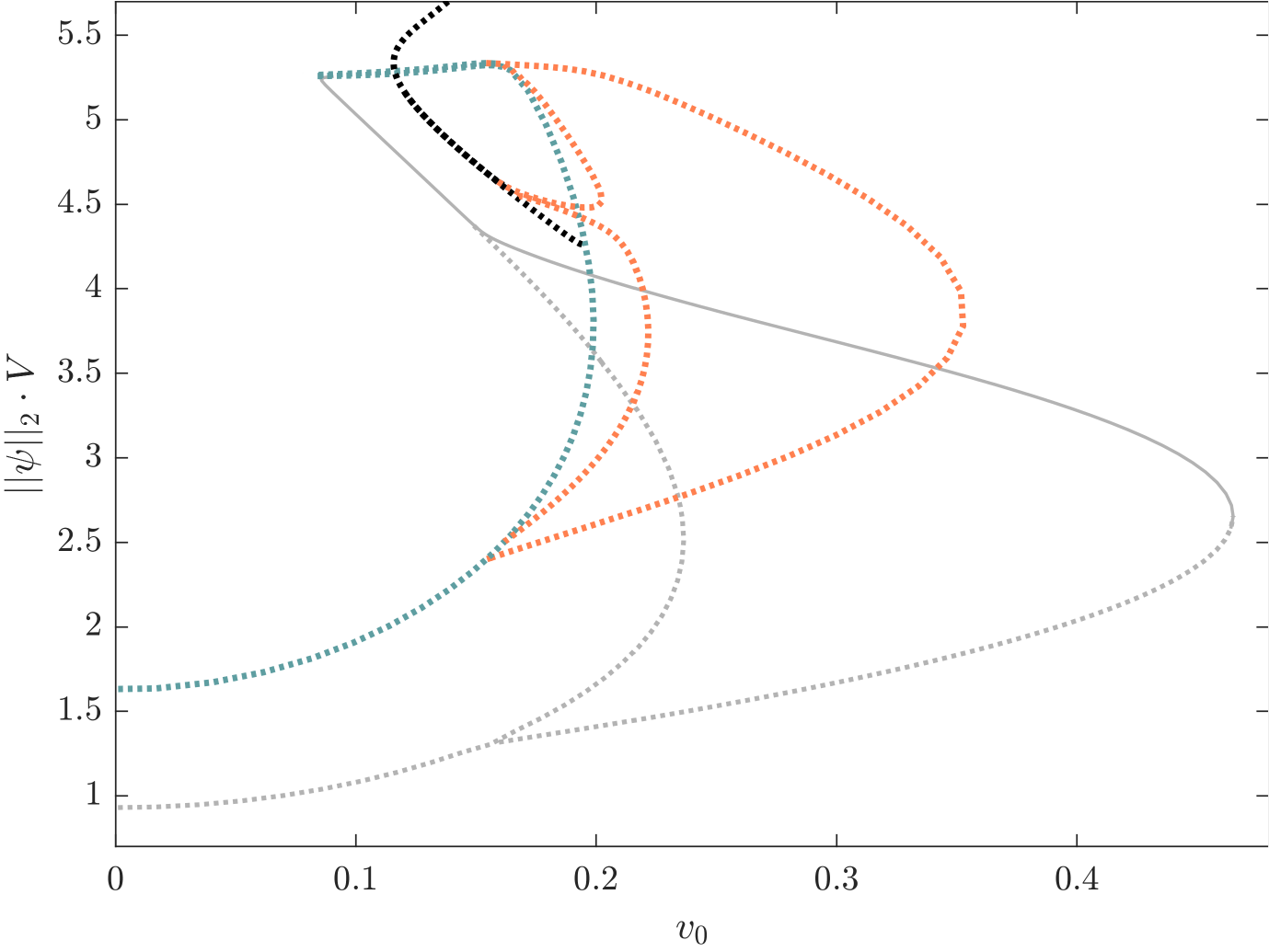}
	 \caption{Summary bifurcation diagram showing $||\psi||_2$ vs $v_0$ for resting and traveling dumbbell-shaped two-peak LS at mean density $\bar{\psi}=-0.8$. Blue and black lines indicate branches of resting symmetric and asymmetric LS, respectively. At $v_0\approx0.15$, states traveling in different directions (orange branches) emerge in various drift bifurcations. See Figs.~\ref{fig:double_parallel} and \ref{fig:double_ortho} for details and selected solution profiles. Thin gray lines correspond to branches of resting and traveling one-peak LS. Remaining line styles, parameters and the domain size are as in Fig.~\ref{fig:singlebump}.}
	 \label{fig:doublebump_all}
	 \end{figure}

	\begin{figure} 
	 \centering\hspace*{-0.0cm}
\begin{overpic}[width=0.5\textwidth]{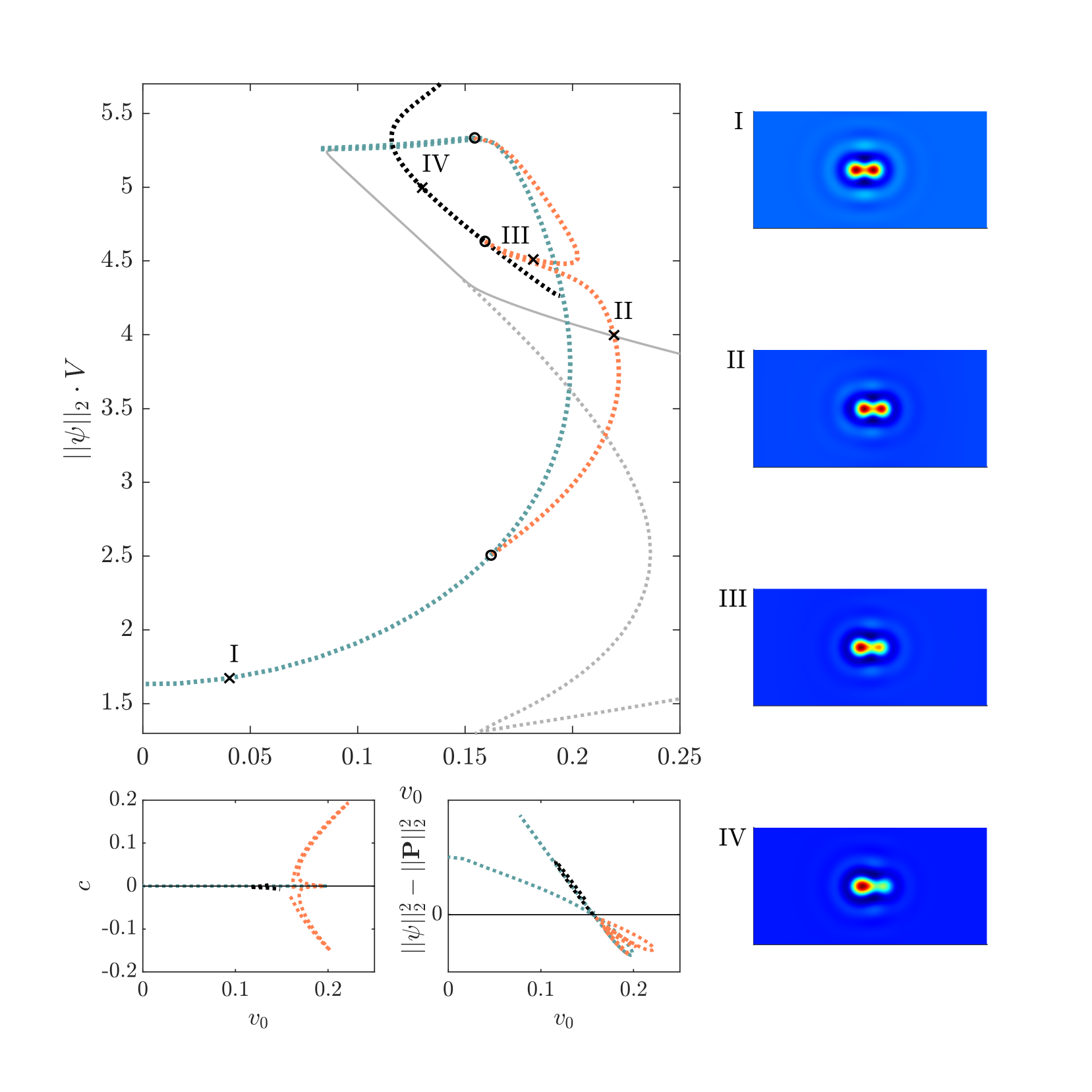}
	 	 \put(82,61.5){\textcolor{white}{\large{$\Rightarrow$}}}
	 	 \put(73,39.5){\textcolor{white}{\large{$\Leftarrow$}}}
	 \end{overpic}
        \caption{(top left) Shown is a subset of the bifurcation curves from Fig.~\ref{fig:doublebump_all}, namely, the  traveling  dumbbell-shaped two-peak LS that move parallel to their long axis, the one-peak states, and the resting two-peak states. The right panels show selected density profiles $\psi(\mathbf{r})$ at points labeled I to IV in the main panel. The resting two-peak LS are destabilized with respect to parallel motion in drift-pitchfork bifurcations at $v_\mathrm{c}\approx0.16$, marked by black circles. On the orange branches of traveling LS, state~III [state~II] travels with the larger [smaller] density peak at the front. The asymmetric steady solution is destabilized in a drift-transcritical bifurcation marked by the circle on the black branch. Here, two branches of TLS with opposite drift velocities emerge. The lower left panels show the drift velocity $c$ as a function of~$v_0$ and the measure $||\psi||_2^2-||\mathbf{P}||_2^2$ that crosses zero at the respective onsets of motion. The remaining line styles, parameters and the domain size are as in Fig.~\ref{fig:doublebump_all}.}
	 \label{fig:double_parallel}
	 \end{figure}
	 
	 \begin{figure}
	 \includegraphics[width=0.5\textwidth]{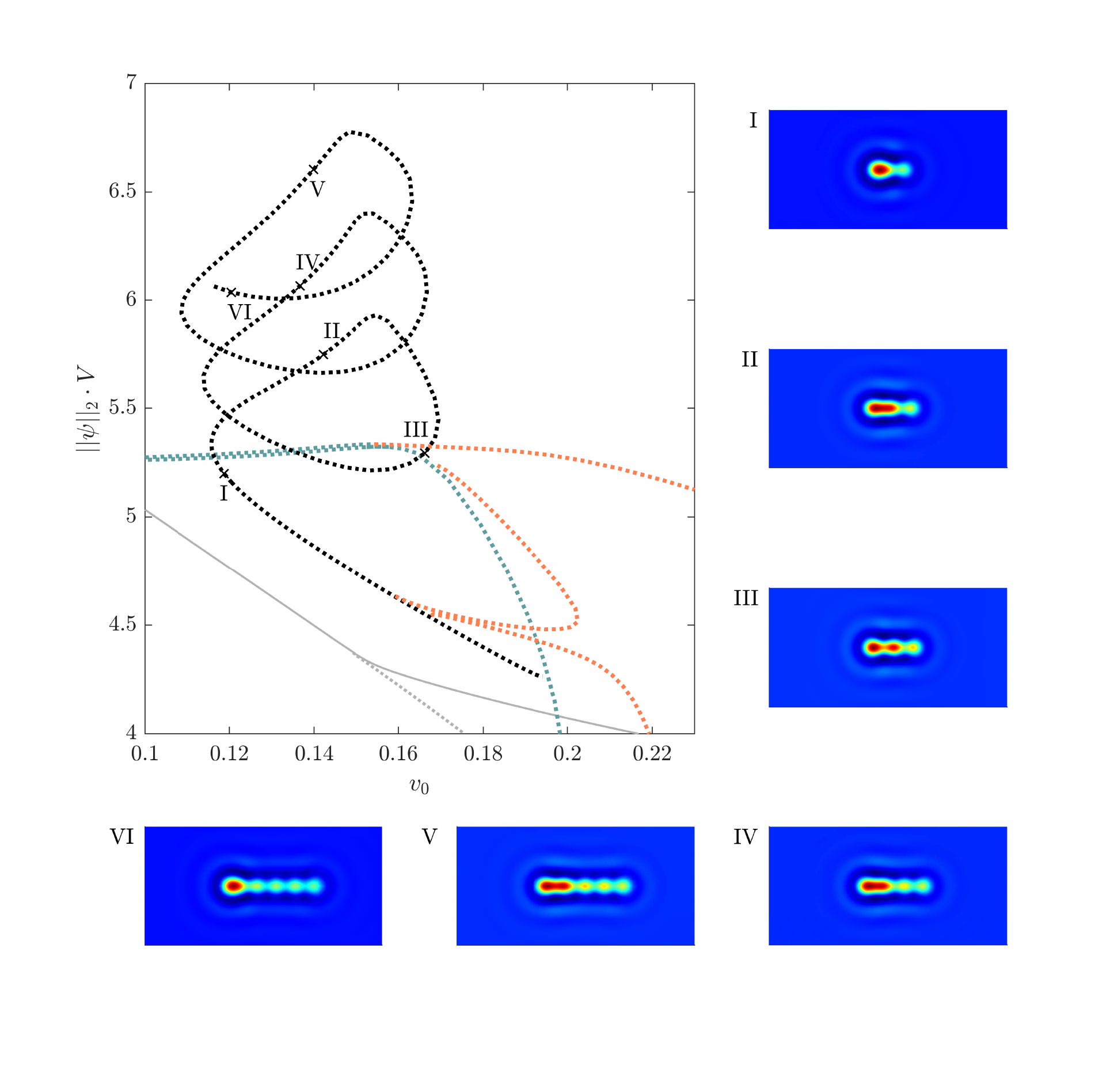}	
	  \caption{Shown is a magnification of Fig.~\ref{fig:double_parallel} completing the branch of steady asymmetric LS (black dashed line). Panels I to VI present selected density profiles $\psi(\mathbf{r})$ at points labeled I to VI in the main panel. The remaining line styles, parameters and the domain size are as in Fig.~\ref{fig:double_parallel}.}
	 \label{fig:assym_snaking}
	 \end{figure}

	 
	\begin{figure} 
	 \centering\hspace*{-0.0cm}
	 \begin{overpic}[width=0.5\textwidth]{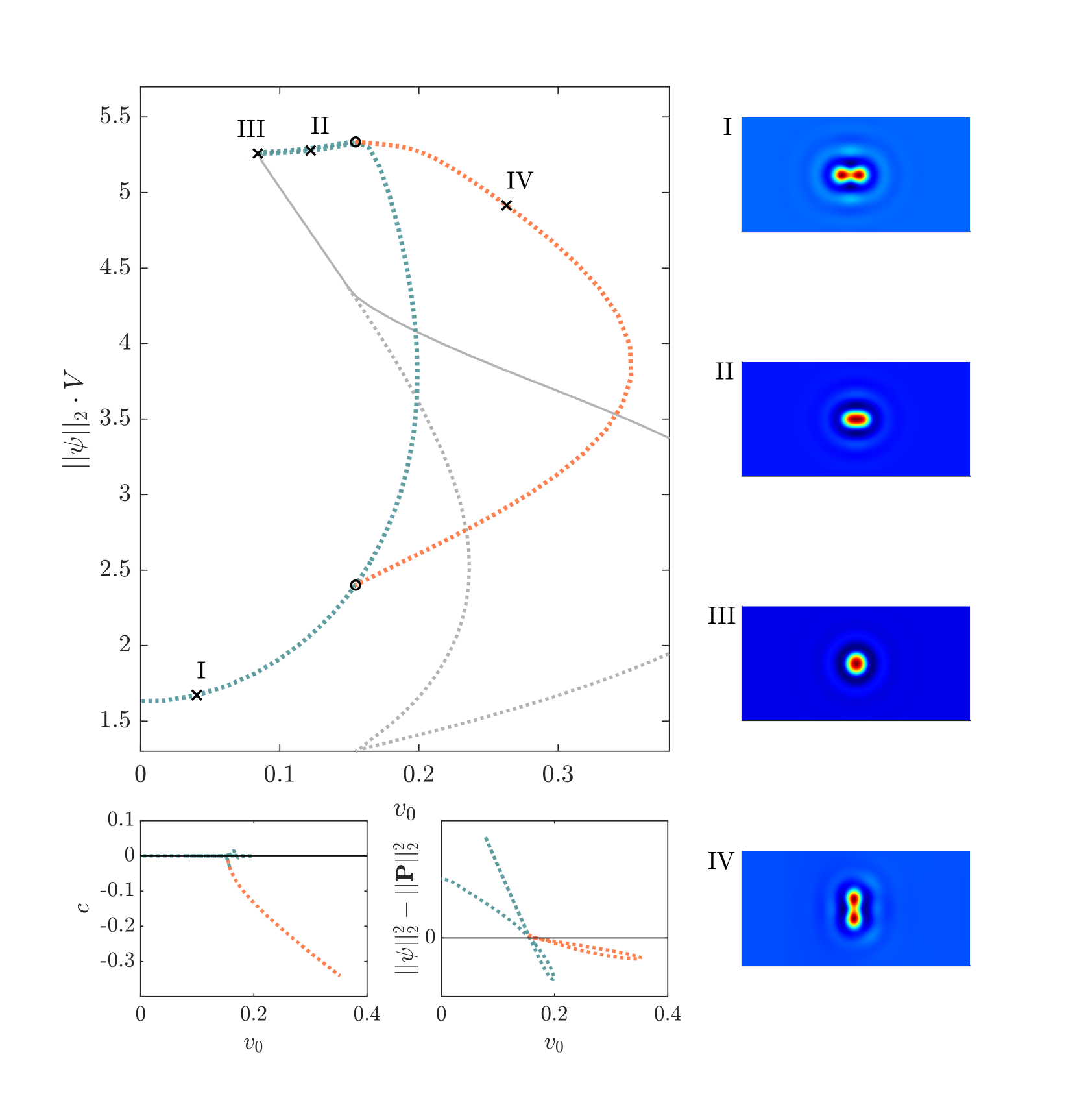}
	 \put(71,18){\textcolor{white}{\large{$\Leftarrow$}}}
	 \end{overpic}
\vspace*{-1.cm}
\caption{(top left) Shown is a subset of bifurcation curves from Fig.~\ref{fig:doublebump_all}, namely, the traveling dumbbell-shaped two-peak LS that move perpendicular to their long axis, the one-peak states, and the resting two-peak states. The remaining panels, line styles, symbols, parameters and the domain size are as in Fig.~\ref{fig:double_parallel}.
}
	 \label{fig:double_ortho}
	 \end{figure}

For $\bar{\psi}=-0.8$, however, the fold of the two-peak LS shifts beyond the threshold for the onset of motion and the two-peak LS also undergo a drift bifurcation. Owing to the additional spatial degree of freedom in 2D a reflection-symmetric structure at rest may undergo motion in two orthogonal directions, longitudinal and transverse, resulting in a drastic change in the overall bifurcation picture. Figure~\ref{fig:doublebump_all} summarizes the intricate bifurcation structure of two-peak crystallites at this value of $\bar{\psi}$. This complicated behavior is disassembled into Figs.~\ref{fig:double_parallel} to \ref{fig:double_ortho} shedding additional light on the different branches of TLS that emerge. 

Figure~\ref{fig:double_parallel} depicts branches of TLS moving longitudinally, i.e., parallel to the long axis of the elongated LS while Fig.~\ref{fig:double_ortho} shows the branches of TLS moving transversely, i.e., parallel to the short axis of the LS. Interestingly, the latter branch extends to higher values of $v_0$.

Figure~\ref{fig:assym_snaking} magnifies the upper part of the bifurcation diagram presented in Fig.~\ref{fig:doublebump_all} completing the branch of asymmetric LS. Interestingly, the branch exhibits tilted snaking like that found in other pattern-forming systems with a conservation law \cite{Knob2016ijam}. However, the behavior does not correspond to the usual snakes-and-ladder structure of snaking branches of symmetric LS connected by branches of asymmetric LS. Instead, the asymmetric LS snake in a slightly slanted, spiralling fashion. With each loop, the asymmetric LS grows in the longitudinal direction by creating one new peak (see, e.g., the density profiles in panels I-VI of Fig.~\ref{fig:assym_snaking}). All the asymmetric states are unstable.

The resting elongated two-peak LS are connected to the rotationally symmetric one-peak solution as indicated in Fig.~\ref{fig:double_ortho}, state III. This point is also a fold near which the stable resting one-peak LS (solid gray line) start to deform into a two-peak LS. Because of the influence of the boundaries this is a continuous transition. In fact, all (reflection-symmetric) resting LS in Fig.~\ref{fig:doublebump_all} correspond to a single branch, similar to the result for the snaking branches as a function of $\bar{\psi}$. The branches of one-peak LS (moving and resting) are shown in light gray with solid (dotted) lines for (un)stable states. Unfortunately, at this value all two-peak LS are still unstable, just as for $\bar{\psi}=-0.9$ (Fig.~\ref{fig:restingdoublebump}).

At $v_\mathrm{c}\approx0.15$, various TLS emerge at drift bifurcations marked in Figs.~\ref{fig:double_parallel} and \ref{fig:double_ortho} by black circles. TLS moving parallel to their long axis (Fig.~\ref{fig:double_parallel}) do not reach activity values as high as the TLS moving transversely (Fig.~\ref{fig:double_ortho}). Figure~\ref{fig:double_parallel} shows that two distinct branches of TLS originate in a drift-transcritical bifurcation on the branch of resting asymmetric states (black). Owing to the lack of left-right symmetry of the density profile, each direction of the drift results in a separate branch. In particular, the TLS on the upper branch move to the left with the larger density peak at their tip (cf. Fig.~\ref{fig:double_parallel}, state III) while the lower branch that emerges (bending towards lower norm) corresponds to TLS with the smaller density peak at the tip (state II). Both branches of TLS terminate on the branch of resting left-right symmetric LS (blue) in respective drift-pitchfork bifurcations (marked by circles). 

We mention that on the scale of Fig.~\ref{fig:double_parallel} the bifurcation occurring on the branch of asymmetric RLS does not exhibit the typical shape of a transcritical bifurcation as both branches of TLS seem to bifurcate towards larger $v_0$. Our identification of this bifurcation as a drift-transcritical bifurcation is based on similar behavior observed in 1D \cite{OphausPRE18} where more precise computations are possible, and for this reason we believe that one of two branches undergoes a fold very close to the transcritical bifurcation. Grid effects make it very hard to remain on branches of RLS and lead to rather blurred onsets of the drift velocity $c$ vs $v_0$. However, with the help of the onset criterion derived in Ref.~\cite{OphausPRE18}, we are able to determine the exact location of all drift bifurcations (lower panels of Fig.~\ref{fig:double_parallel}).

Note that the bifurcation diagrams presented in Figs.~\ref{fig:doublebump_all} and \ref{fig:double_parallel} are incomplete and focus only on the transitions to TLS. In particular, further evolution of the RLS branch for higher values of the $L_2$ norm is omitted here and will be the subject of further research.

Figure~\ref{fig:double_ortho} explores the branches of traveling two-peak LS that travel transversely. Here, the picture is simpler. As for the traveling one-peak LS, a branch of TLS stretches between a pair of drift bifurcations highlighted by black circles. Panels on the right show selected density profiles.   

\begin{figure}[t]
	\centering
	 \includegraphics[width=0.5\textwidth]{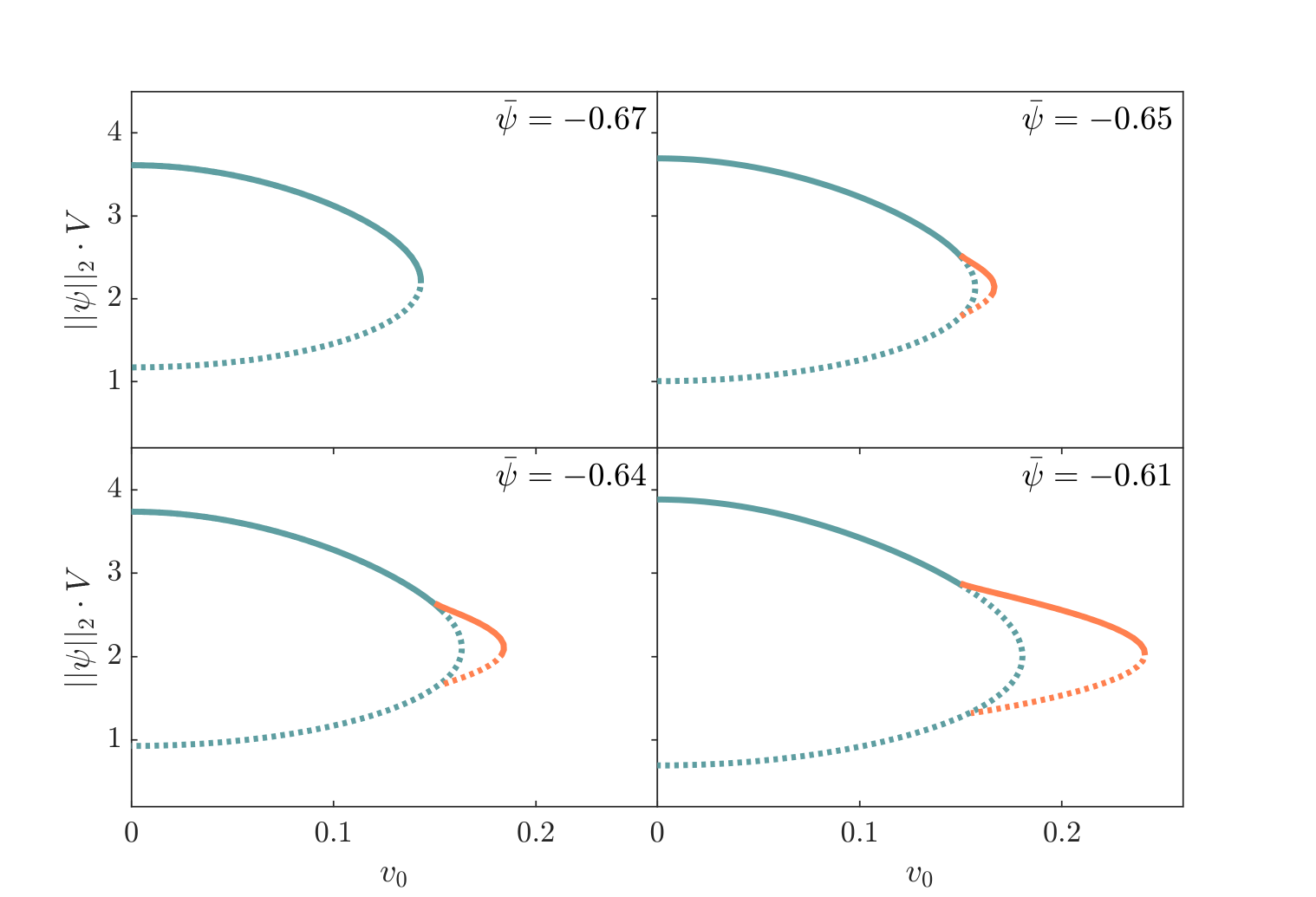}
         \caption{A sequence of bifurcation diagrams $||\psi||_2$ vs $v_0$ showing how traveling one-peak LS (orange line) come into existence with increasing mean density $\bar{\psi}$ (from top left to bottom right) at $\epsilon=-0.98$ (corresponding to the value used in \cite{MenzelLoewen}). Two drift-pitchfork bifurcations appear simultaneously at the saddle-node bifurcation of the branch of resting LS (blue line). Increasing $\bar{\psi}$ further expands the range of existence of traveling LS toward larger $v_0$ and the drift-pitchfork bifurcations separate. The onset of motion is always at $v_\mathrm{c}\approx 0.15$. Remaining line styles, parameters and the domain size are as in Fig.~\ref{fig:singlebump}.}
 \label{fig:onset_of_motion_4psibar}
 \end{figure}

Overall, the bifurcation structure of traveling two-peak LS is much more intricate than that of the single-peak LS. Moreover, increasing $\bar{\psi}$ from $-0.9$ to $-0.8$ drastically changes the bifurcation structure. Figure~\ref{fig:onset_of_motion_4psibar} illustrates how the drift bifurcations and TLS come into existence by showing a series of four bifurcation diagrams for increasing values of mean density $\bar{\psi}$. Between $\bar{\psi}=-0.67$ and $-0.65$ a pair of drift bifurcations is created. Their origin coincides with the fold of the branch of resting states (blue lines) as shown by a two-parameter continuation. For increasing $\bar{\psi}$, the region of existence of TLS grows as the fold of the TLS branch moves to higher values of $v_0$ while the threshold activity for the onset of migration, $v_\mathrm{c}$, stays practically constant. These results are consistent with the results of extensive fold continuation in 1D. Note that for Fig.~\ref{fig:onset_of_motion_4psibar} we have used $\epsilon=-0.98$ in contrast to previous figures with $\epsilon=-1.5$. 

With the various TLS obtained by continuation in $v_0$ in hand, we are now able to construct a morphological phase diagram (next section). This is followed by an examination of the bifurcation diagrams for fixed $v_0$ as the mean density $\bar{\psi}$ varies and a study of the phenomenon of slanted homoclinic snaking for active crystallites at $v_0>0$ in Sec.~\ref{sec:activesnaking}.

\subsection{\label{sec:phasediagram}Morphological phase diagram}
	 
	\begin{figure}	
	\centering
	\hspace*{0cm}
    \includegraphics[width=0.48\textwidth]{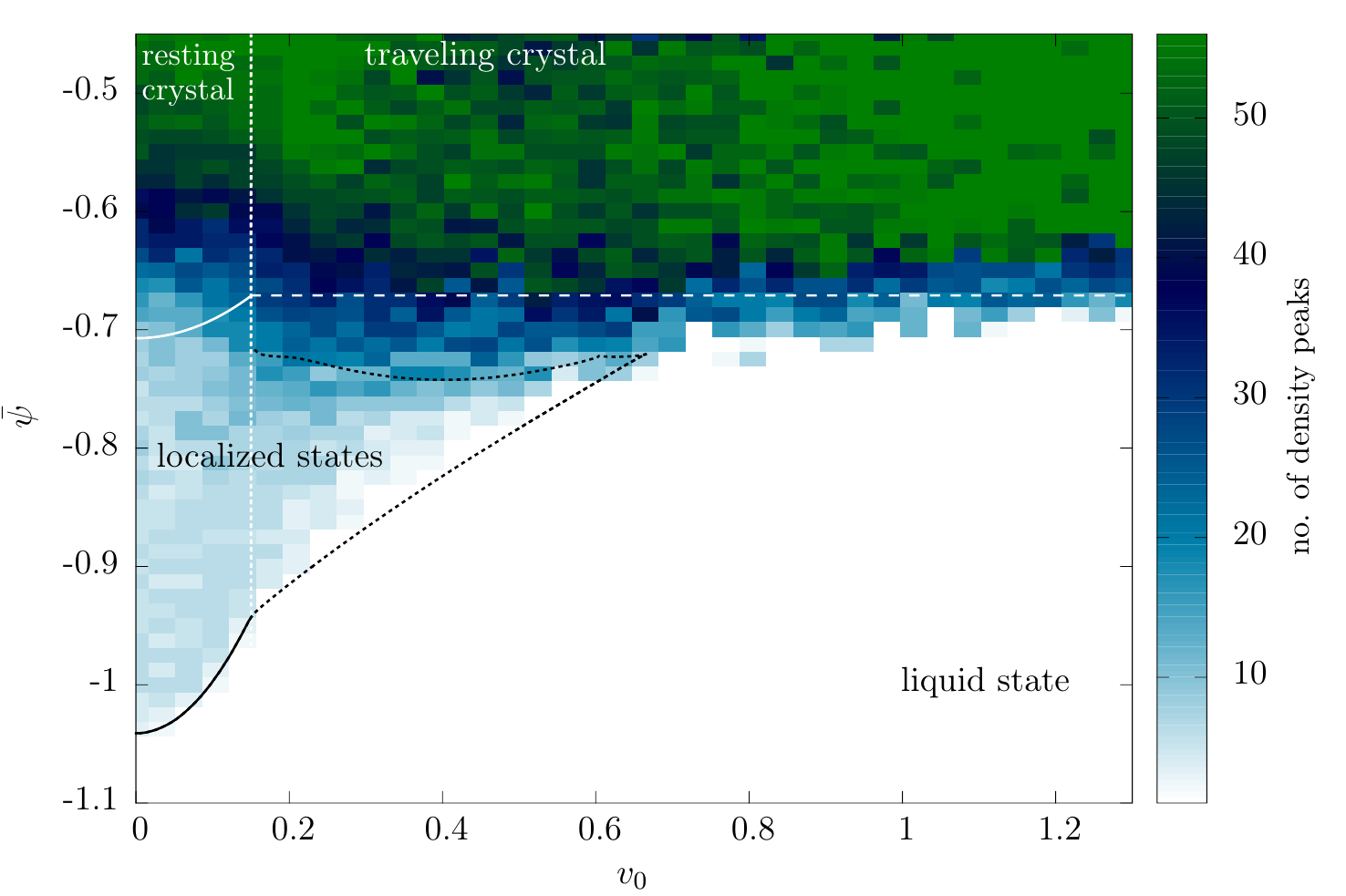}
	 \caption{
           Morphological phase diagram for the aPFC model in 2D in the plane spanned by the activity $v_0$ and the mean density $\bar{\psi}$ as obtained through systematic time simulations. The region of stable liquid state is white, while crystalline structures of various size exist in the colored areas. The color bar indicates the number of density peaks formed in the domain of size $8L_\mathrm{c} \times 7L_\mathrm{a}$ with $L_\mathrm{a}=2L_\mathrm{c}/\sqrt{3}$ and $L_\mathrm{c}=2\pi$. Regions where resting and traveling LS exist are marked by shades of blue while domain-filling periodic patterns are shown as green (56 peaks). The various lines in the diagram, the initial conditions for the simulations, and the peak counting procedure are described in the text. 
           The remaining parameters are $\epsilon=-1.5$, $C_1=0.1$, $C_2=0$ and $D_\mathrm{r}=0.5$ as used throughout Sec.~\ref{sec:v0}. The parameter increments between simulations are $\Delta v_0 = 0.035$ and $\Delta\bar{\psi}=0.0125$. See Fig.~\ref{fig:phasediagram_profiles} for a magnification of the region close to the onset of motion and selected density profiles.}
	 \label{fig:phasediagram}
	 \end{figure}
	 
	\begin{figure}	
	\centering
%
   \begin{overpic}[width=0.5\textwidth]{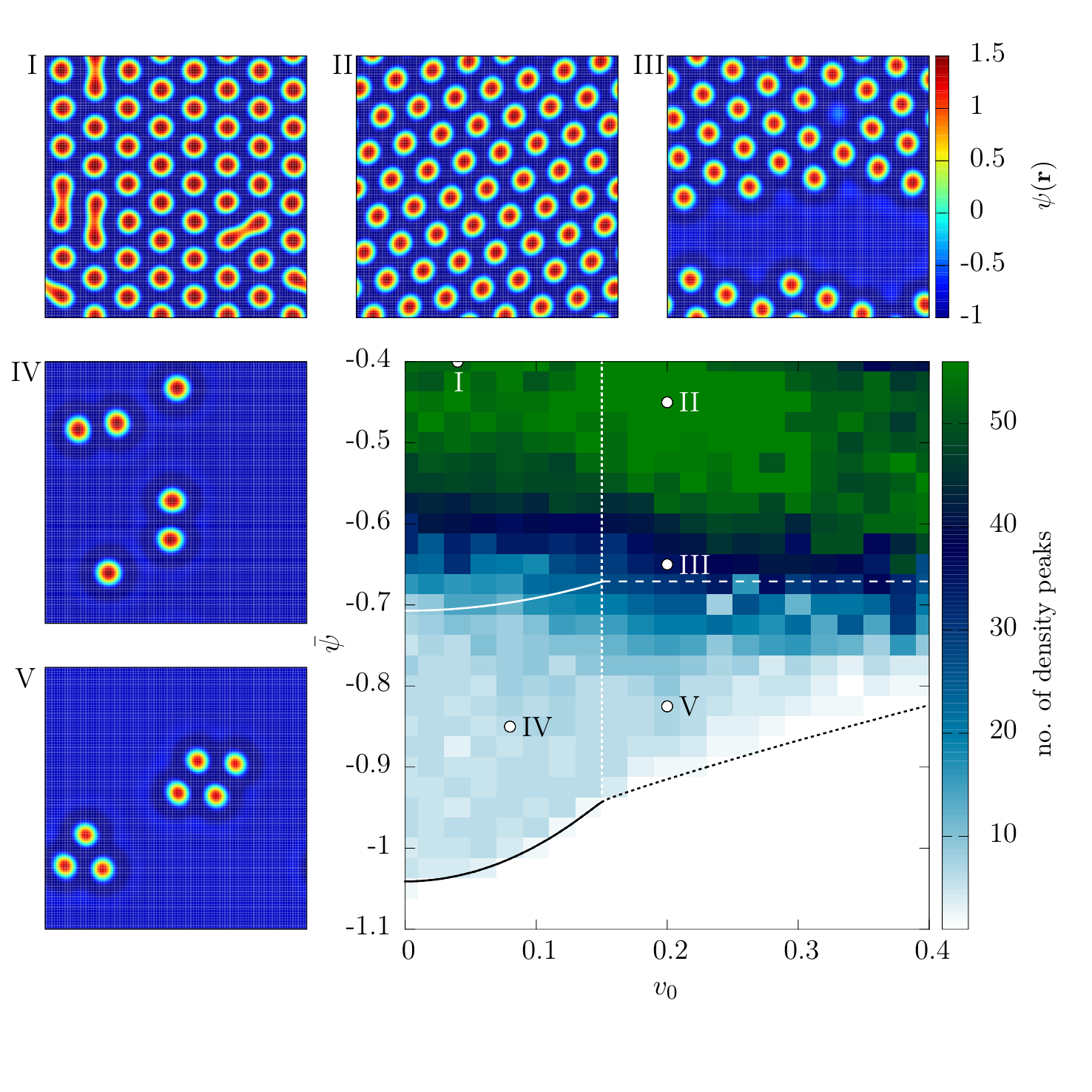}
	\put(49,79.5){\rotatebox{-20.}{\makebox(0,0){\strut{}\textcolor{white}{\normalsize{$\Longrightarrow$}}}}}%
	\put(68.3,84.7){\rotatebox{10.}{\makebox(0,0){\strut{}\textcolor{white}{\normalsize{$\Longrightarrow$}}}}}%
	\put(23.5,31){\rotatebox{27.}{\makebox(0,0){\strut{}\textcolor{white}{\normalsize{$\Longrightarrow$}}}}}%

	\end{overpic} 
	 \vspace*{-1.7cm}

	 \caption{The large panel shows a magnification of the region close to the onset of motion in the morphological phase diagram in Fig.~\ref{fig:phasediagram}. The parameter increments between simulations are $\Delta v_0 = 0.02$ and $\Delta\bar{\psi}=0.025$. 
The small panels show selected density profiles $\psi(\mathbf{r})$ at points labeled I to V in the phase diagram after the time simulations have converged. Arrows indicate direction of motion. Shown are (I) resting hexagonal pattern close to the transition to stripes, (II) traveling hexagonal pattern, (III) traveling cluster of hexagonal order, (IV) resting LS, and (V) traveling LS.}
	 \label{fig:phasediagram_profiles}
	 \end{figure}

Before discussing in detail the bifurcation structure as a function of $\bar{\psi}$ and changing the tempera\-ture-like parameter $\epsilon$ to allow for snaking, we conclude the discussion of active crystallites at $\epsilon=-1.5$ by presenting a large-scale morphological phase diagram in the parameter plane spanned by $v_0$ and $\bar{\psi}$. The phase diagram is determined numerically by counting peaks of $\psi(\mathbf{r},t)$ after a sufficiently long transient. To favor the creation of LS, six density bumps are superposed at random positions on the homogeneous phase marginally perturbed by white noise. For the polarization $\mathbf{P}$, we choose a randomly perturbed trivial state $\mathbf{P}_0=\mathbf{0}$ as initial condition.

The domain with periodic boundaries has a size of $8L_\mathrm{c} \times 7L_\mathrm{a}$ with $L_\mathrm{a}=4\pi/\sqrt{3}$ being the side length of a hexagon/triangle and $L_\mathrm{c}=2\pi$ the critical wavelength at the onset of crystallization. As already mentioned, $L_\mathrm{c}$ is the height of triangles found in the hexagonal pattern. This domain size results in a maximum number of 56 density peaks in a periodic array [cf.~Fig.~\ref{fig:phasediagram_profiles}(I)].

In Fig.~\ref{fig:phasediagram} periodic states with around 56 density peaks are displayed in green, whereas LS exist within the blueish area. The white area without any density peaks corresponds to the liquid state $\psi(\mathbf{r})=0$. The white lines indicate the stability limits obtained from linear stability of the liquid phase, with the vertical white line indicating the onset of motion at $v_\mathrm{c}$ ($v_{\mathrm{c}}\approx0.15$, independently of $\bar{\psi}$). The limits of the existence of LS are determined by a two-parameter continuation of their fold. The black lines show the position of folds of resting one-peak LS (solid black) and of traveling one-peak LS (dotted black). The position of the saddle-node bifurcations of 2D TLS starts to shift backwards, towards smaller values of $v_0$ at $v_0\approx0.7$. This is a major difference from the one-peak TLS in one spatial dimension which exist to arbitrarily high $v_0$.

The time simulations indicate large areas of existence of various active LS (blueish area). The extent of the LS region ranges from single density peaks (light blue) to patches of LS almost filling the entire domain (dark blue). Selected solution profiles $\psi(\mathbf{r})$ can be found in Fig.~\ref{fig:phasediagram_profiles}. The phase diagram also illustrates how hexagonal periodic states change their shape towards a stripe pattern resulting in a lower number of density peaks [Fig.~\ref{fig:phasediagram_profiles}(I)]. For such patterns each elongated ridge is only counted as a single density peak. Close to the limit of linear stability of the liquid state, large patches of localized crystalline order coexist with the uniform state [Fig.~\ref{fig:phasediagram_profiles}(III)]. Resting LS ($v_0<0.15$, left of the vertical dotted line denoting the onset of motion) exist down to low values of $\bar{\psi}\approx-1.05$. Increasing activity melts most of these LS ($\bar{\psi}<-0.95$) and the $v_0$-range of their existence contracts as $\bar{\psi}$ decreases. Higher $\bar{\psi}$ favors traveling LS ($-0.95\lesssim \bar{\psi}\lesssim -0.7$, $0.15\lesssim v_0\lesssim 0.7$) and traveling crystals that fill the entire domain are present for $\bar{\psi}\gtrsim-0.65$, the linear stability threshold of the liquid state, and $v_0\gtrsim0.15$, where the blue region terminates giving way to green areas.

We also see that traveling periodic states exist to arbitrarily high activities and do not melt, unlike most LS, in agreement with similar observation in 1D \cite{OphausPRE18}.

\subsection{\label{sec:activesnaking}Snaking of active crystallites}

\begin{figure}
\centering
\includegraphics[width=0.5\textwidth]{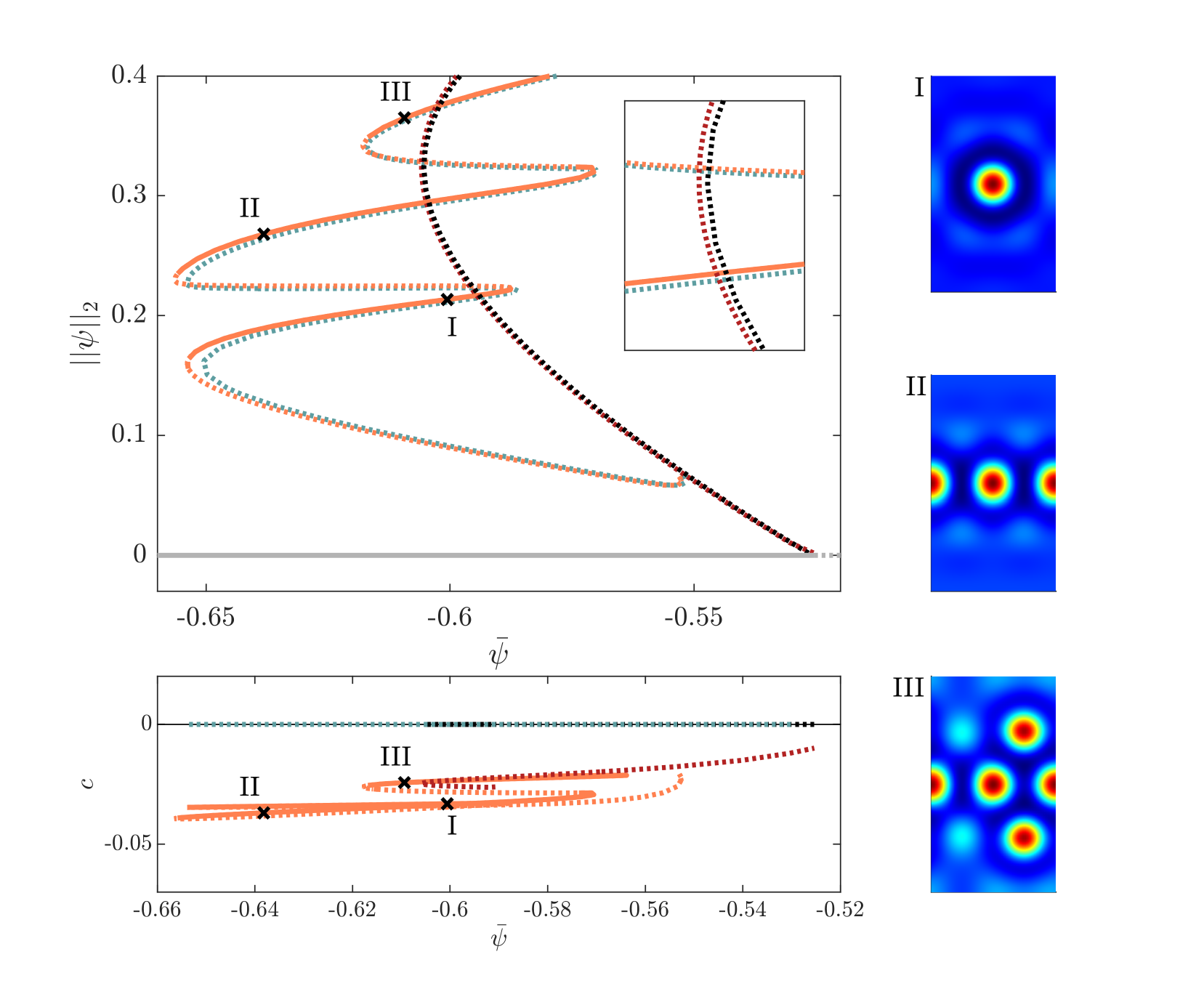}
 \caption{Bifurcation diagram showing slanted snaking of resting and traveling LS at $\epsilon=-0.98$. Shown is the norm $||\psi||_2$ as a function of the mean density $\bar{\psi}$ with activity fixed at the still relatively low value $v_0=0.151>v_\mathrm{c}$ that allows for the coexistence of resting and traveling states. Labels I to IV mark the location of the stable traveling LS shown on the right. The liquid phase (gray line) with norm zero is destabilized at $\bar{\psi} \approx -0.53$ and a branch of traveling periodic patterns (dark red) emerges. Close to the first primary bifurcation, a branch of resting crystals (black) emerges in another primary bifurcation. Resting and traveling LS (blue and orange), respectively, bifurcate in secondary bifurcations from these branches. Since $v_0>v_\mathrm{c}$, all resting solutions are unstable. The inset illustrates the small separation of the branches in terms of their norm. The lower panel shows the drift velocity $c$ as a function of $\bar{\psi}$. Since $c<0$, all TLS move to the left. The domain size is $2L_\mathrm{a} \times 4L_\mathrm{c}$ while the remaining line styles and parameters are as in Fig.~\ref{fig:singlebump}.}
\label{fig:activesnaking0151}
 \end{figure}

\begin{figure}	
\centering

\includegraphics[width=0.5\textwidth]{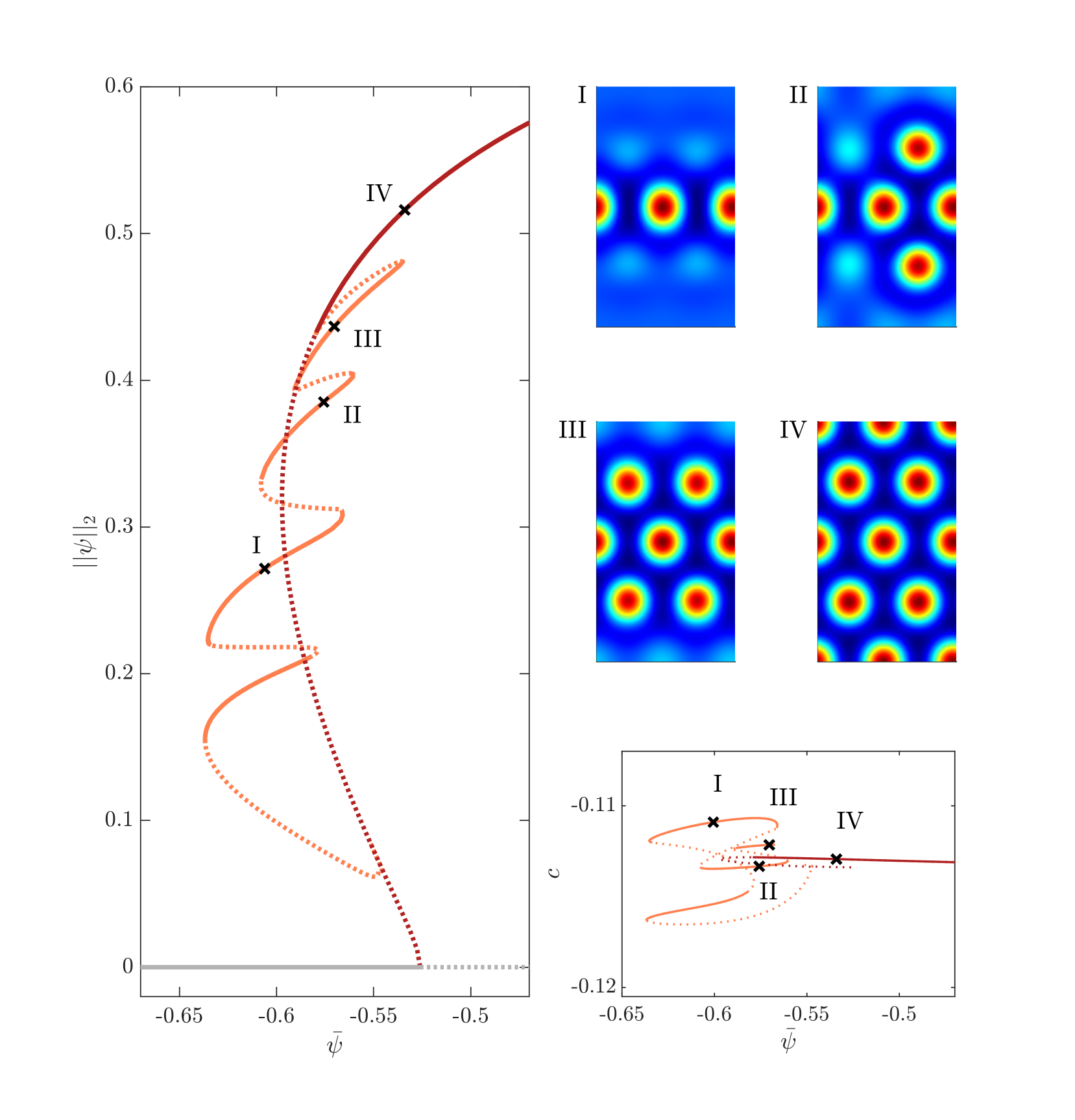}
\vspace*{-0.7cm}
\caption{Bifurcation diagram $||\psi||_2$ vs $\bar{\psi}$ for traveling hexagonal patterns and traveling LS at the relatively high activity $v_0=0.18$ where no resting states exist. Labels I to IV denote the locations of the stable traveling LS shown on the right. The bottom right panel shows the drift velocity $c$ as a function of $\bar{\psi}$. All states travel to the left. Domain size, line styles and the remaining parameters are as in Fig.~\ref{fig:activesnaking0151}.}
\label{fig:activesnaking018}
 \end{figure}

In this section, we explore in detail the bifurcation structure of both resting and traveling \textit{active} crystallites as a function of the mean density $\bar{\psi}$. Having analyzed the slanted snaking of passive LS (Sec.~\ref{sec:passive_snaking}), we now wish to examine the influence of the activity parameter $v_0$ on the snaking of 2D LS and the response of 2D TLS to varying $\bar{\psi}$.

The value $\epsilon=-1.5$ of the effective temperature turns out to be too low to support continuous snaking of both passive and active LS (cf. Fig.~\ref{fig:no_snaking_eps15}). At these values of $\epsilon$ the snaking branches most likely break up into disconnected pieces. For this reason we increase the temperature to $\epsilon=-0.98$ as done for passive crystallites in Sec.~\ref{sec:passive_snaking}. In addition, this value is also employed in \cite{MenzelLoewen} where the aPFC model was introduced. There, diffusion is set to $C_1=0.2$ leading to $v_\mathrm{c}\approx0.3$. The high diffusion causes many crystallites to melt before motion can set it. We therefore stick to $C_1=0.1$ as used in the previous sections, for which the threshold for the onset of migration is $v_\mathrm{c}\approx0.15$.

Figure~\ref{fig:activesnaking0151} depicts the bifurcation diagram at $v_0=0.151$, slightly above $v_\mathrm{c}$ and allowing for TLS. The overall picture is similar to the slanted homoclinic snaking found for passive LS. The branches of LS bifurcate from periodic solutions that emerge in subcritical primary bifurcations from the destabilized liquid state and extend well below the folds of the periodic state. Besides the resting crystal (Fig.~\ref{fig:activesnaking0151}, dotted black branch), there is a branch of traveling spatially extended patterns (red). Both crystals are of hexagonal order and their norms differ only slightly. The inset in Fig.~\ref{fig:activesnaking0151} enlarges the region close to the folds of the periodic states, illustrating their small separation. 

The resting and traveling LS branch off from the resting and traveling periodic solutions in secondary bifurcations at small amplitude. Since the value of the activity parameter $v_0$ is above the threshold for migration, all RLS (blue branch) are unstable as indicated by dotted lines. The TLS exhibit the typical alternation of stable and unstable states familiar from slanted snaking of passive LS. Like the branches of periodic solutions the resting and traveling LS have very similar $L^2$-norm. 

The lower panel of Fig.~\ref{fig:activesnaking0151} shows the drift velocity $c$ of the respective solutions as a function of $\bar{\psi}$. Evidently, RLS (blue) and the resting crystal (black) have $c=0$. The velocity of the traveling crystal (red) rises slightly as the crystal grows. In contrast to the main panel, here it is easy to see how the TLS branch off from the traveling periodic solution. All in all, the drift velocity $c$ does not depend strongly on $\bar{\psi}$. 

The same holds at $v_0=0.18$ as shown in Fig.~\ref{fig:activesnaking018}. The bottom right panel reveals that $c$ is almost independent of the mean density $\bar{\psi}$. Because $v_0=0.18$ is beyond the positions of the folds of RLS, resting solutions are no longer present. As also observed at smaller $v_0$, the TLS (orange) emerge in a secondary bifurcation from the branch of traveling crystals (red). With increasing $\bar{\psi}$, the TLS grow by adding density peaks until the whole domain is filled by the crystalline state and the TLS branch terminates on the branch of traveling periodic states [cf.~Fig~\ref{fig:activesnaking018}(IV)]. In contrast to passive snaking, the growth of the TLS does not occur by adding density peaks layer by layer. The broken symmetry at $v_0>v_\mathrm{c}$ seems to favor growth via the addition of pairs of density peaks, maintaining reflection symmetry with respect to $y=0$ at all times, as shown in panels (I) and (II). The different growth pattern is reflected in the larger number of undulations of the snaking branch as compared to the passive case in Fig.~\ref{fig:rect_snaking_eps098}.

Overall, we find that the mean density $\bar{\psi}$ does not have a strong influence on the drift velocity $c$ of the traveling states. In addition, there are no connections between the branches of RLS (of hexagonal order) and TLS (also of approximately hexagonal order) as $\bar{\psi}$ varies. Hence, changes in the mean density cannot directly induce drift instability (although a suitable $\bar{\psi}$ is necessary for drift instabilities to occur when varying $v_0$). And rather unexpectedly (when taking the 1D results~\cite{OphausPRE18} as a guide), the branch of TLS at $\epsilon=-0.98$ exhibits slanted homoclinic snaking much as observed for RLS in the passive system and in active ones at small $v_0$. We do not expect the spatial dimension to play a role here; more likely, the presence of slanted snaking is solely a consequence of choosing an effective temperature $\epsilon$ that is not too negative. We mention that slanted snaking associated with traveling structures is present even in nonconserved systems~\cite{arik}, likely a consequence of the fact that the drift speed is itself a nonlocal property.

\section{\label{crystals}Periodic States}
In two spatial dimensions, different periodic patterns can be distinguished. Besides stripes that can be regarded as a 2D extension of the periodic states determined in 1D~\cite{OphausPRE18}, the aPFC model exhibits both hexagonal and rhombic structures. In this section, we analyze the periodic states that emerge in the 2D aPFC model and, in particular, study their bifurcation structure.

\subsection{Crystal structure and activity}

	\begin{figure}
\begin{overpic}[width=0.5\textwidth]{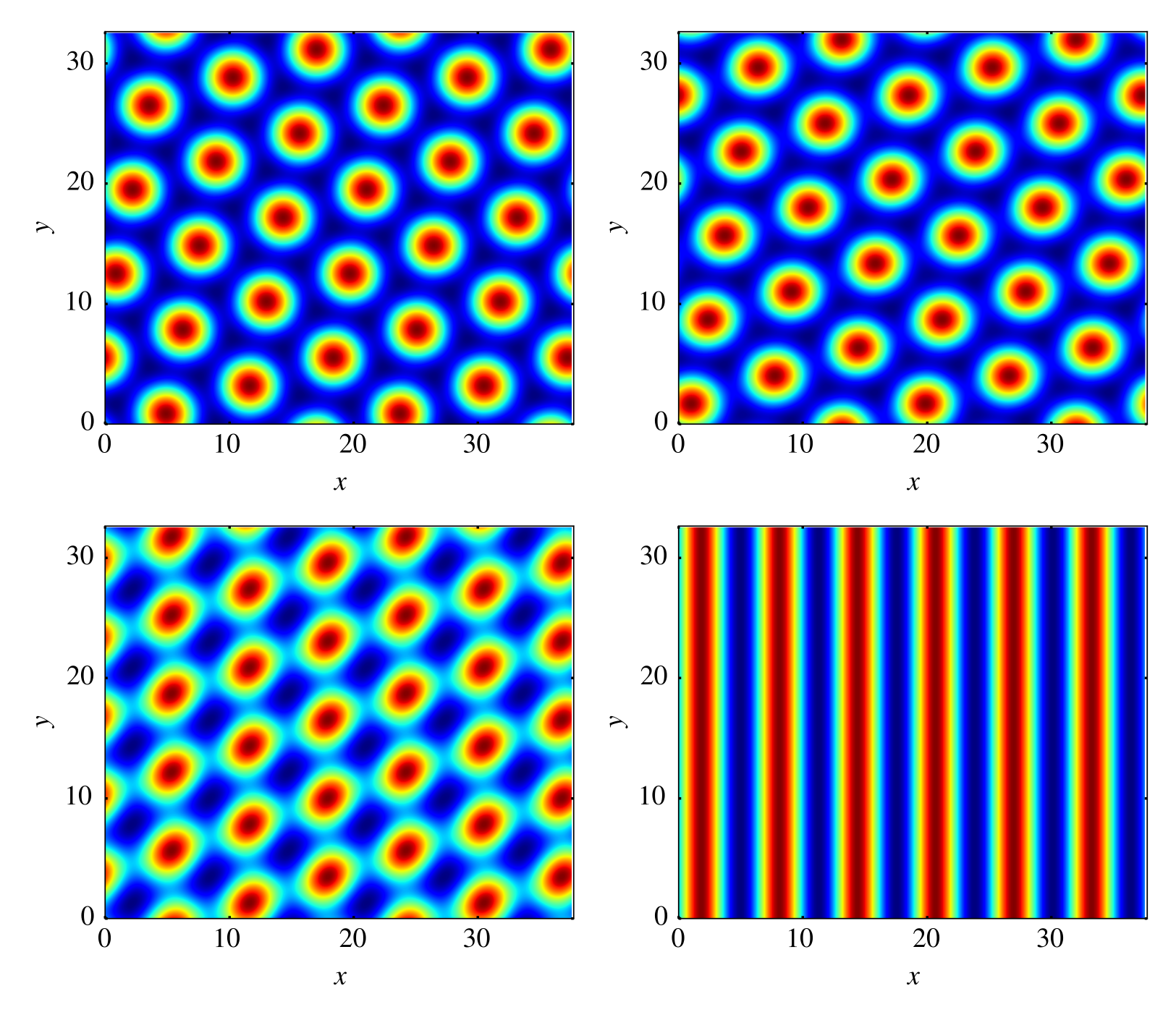}
	  \put(3,84){(a)}
	  \put(51.5,84){(b)}
	  \put(3,42){(c)}
	  \put(51.5,42){(d)}
	  \put(80,67.){\rotatebox{-160.}{\makebox(0,0){\strut{}\textcolor{white}{\Huge{$\Longrightarrow$}}}}}%
	  \put(33,16.5){\rotatebox{-215}{\makebox(0,0){\strut{}\textcolor{white}{\Huge{$\Longrightarrow$}}}}}%
	  \put(64,16){\rotatebox{0}{\makebox(0,0){\strut{}\textcolor{white}{\Huge{$\Longrightarrow$}}}}}%
	  \end{overpic} 
          \caption{Selected snapshots of periodic density patterns $\psi(\mathbf{r})$ at the fixed mean density $\bar{\psi}=-0.4$ as obtained for increasing activity by time simulation. Panel (a) shows a resting hexagonal pattern at $v_0=0.25$, (b) a traveling hexagonal pattern at $v_0=0.3$, (c) a traveling rhombic pattern at $v_0=0.8$, and (d) a traveling stripe pattern at $v_0=1.5$. 
            The respective directions of motion are indicated by white arrows. The domain size is $ 6 L_\mathrm{c} \times \, 5 L_\mathrm{a}$ while the remaining parameters are $\epsilon=-0.98$, $C_1=0.2$, $C_2=0$ as in Ref.~\cite{MenzelLoewen}.}
	 \label{fig:DNS_ML}
	 \end{figure}
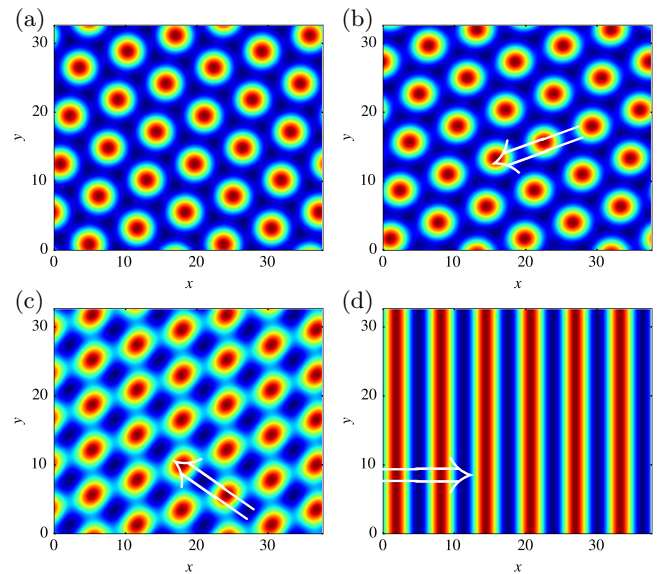
	 
The original passive PFC model exhibits crystalline hexagonal patterns in certain ranges of the temperature $\epsilon$ and mean density $\bar{\psi}$. Changing the mean density can lead to transitions to stripes \cite{TARG2013pre}. These transitions can also be induced by the activity $v_0$. In the original paper \cite{MenzelLoewen} introducing the aPFC model, numerical time simulations show a transition from a resting hexagonal pattern to traveling hexagons with increasing $v_0$. A further increase leads to a transition to traveling rhombic patterns and, ultimately, to traveling stripes. Snapshots from time simulations at certain values of $v_0$ and the same set of control parameters as in \cite{MenzelLoewen} are shown in Fig.~\ref{fig:DNS_ML} and these reproduce previously made observations. 

The domain is of size $ 6 L_\mathrm{c} \times \, 5 L_\mathrm{a}$ with critical wavelength $L_{\mathrm{c}}=2\pi$ and side length $L_\mathrm{a}=4\pi/\sqrt{3}$ accounting for 30 density peaks in hexagonal order. At $v_0=0$, resting hexagons are oriented parallel to the $y$-axis and perfectly match the aspect ratio of $L_x$ and $L_y$. As $v_0$ is increased the wave vector and geometry of the pattern change. The whole crystalline structure reacts by a rotation within the periodic domain [cf.~Fig.~\ref{fig:DNS_ML}(a) and (b)] thereby adjusting its position such that the dominant wave vectors fit into the domain. 

Rhombic [Fig.~\ref{fig:DNS_ML}(c)] and stripe patterns  [Fig.~\ref{fig:DNS_ML}(d)] orient themselves parallel to the $y$-axis as $L_x$ is a multiple of $L_\mathrm{c}=2\pi$. Following the drift instability at $v_\mathrm{c}\approx 0.3$, these patterns travel with a constant speed $c$ while keeping their spatial periodicity. White arrows indicate the direction of motion. Stripes always travel perpendicular to their orientation. Hexagons and rhombi also exhibit specific directions of motion. Therefore, the patterns have to be correctly oriented when employing numerical continuation with the particular boundary conditions discussed in Sec.~\ref{sec:num2d}. These only permit motion in the $x$-direction.

\subsection{Pattern selection and bifurcation structure}

\begin{figure} \vspace*{0.cm}
	 \centering\hspace*{.0cm}
	 \includegraphics[width=0.5\textwidth]{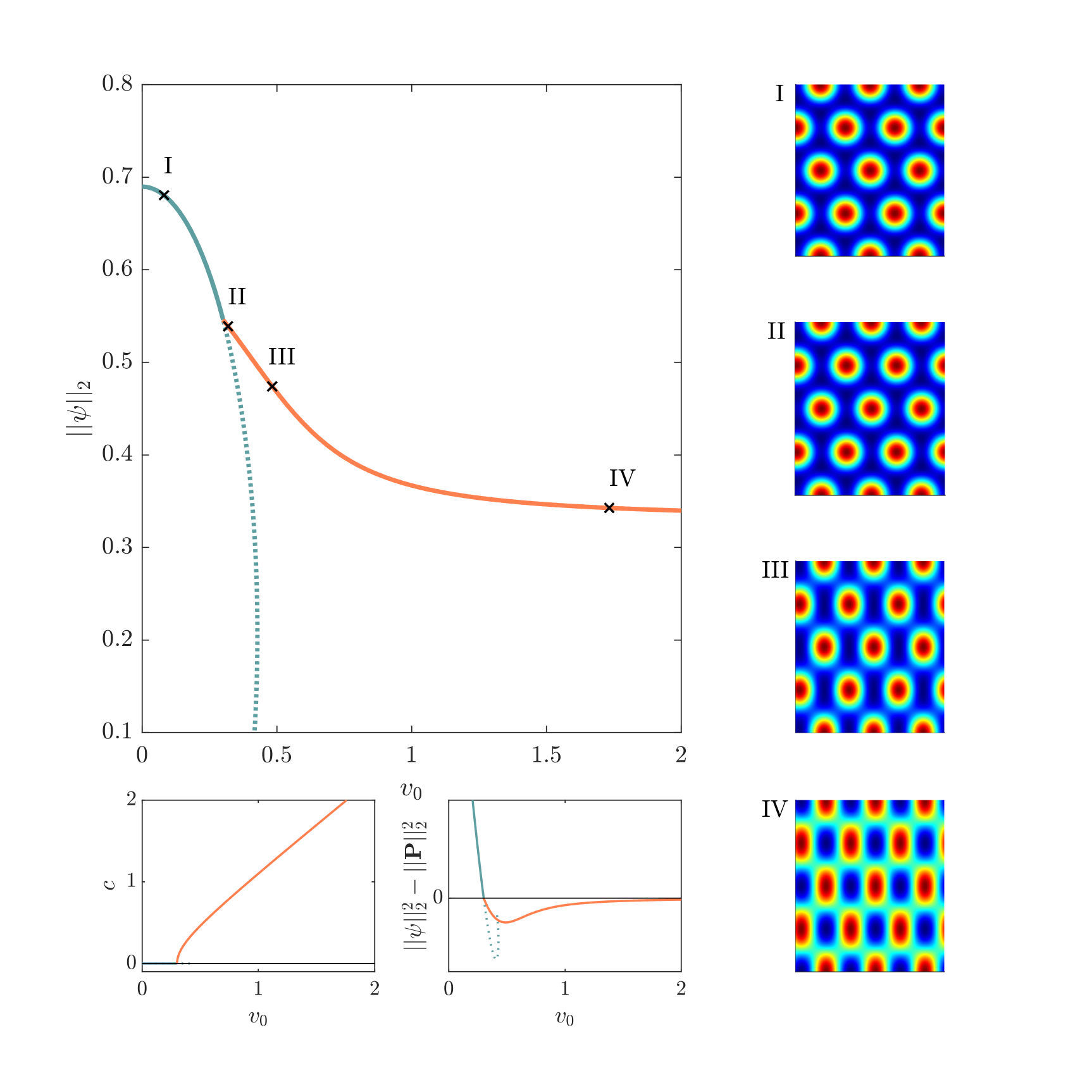}
	 \caption{The main panel shows the bifurcation diagram $||\psi||_2$ vs $v_0$ for hexagonal patterns oriented such that the onset of motion is parallel to an edge. Resting hexagons (blue line) are stable (solid line) until a drift-pitchfork bifurcation occurs at $v_\mathrm{c}\approx0.3$ where a branch of stable traveling hexagonal patterns (orange line) emerges. The corresponding drift velocity $c$ is shown in the lower left panel while the lower right panel shows the measure $||\psi||_2^2-||\mathbf{P}||_2^2$ that crosses zero at the drift bifurcation. On the right selected solution profiles $\psi(\mathbf{r})$ at the locations labeled I to IV in the bifurcation diagram are shown. Profiles II-IV travel in the $x$-direction to the right. The domain size is $V=3 L_\mathrm{a} \times\,4L_\mathrm{c}$ and the remaining parameters are as in Fig.~\ref{fig:DNS_ML}.}
	 \label{fig:cont_hex}
\end{figure}

\begin{figure}[t]
	 \centering
	  \includegraphics[width=0.5\textwidth]{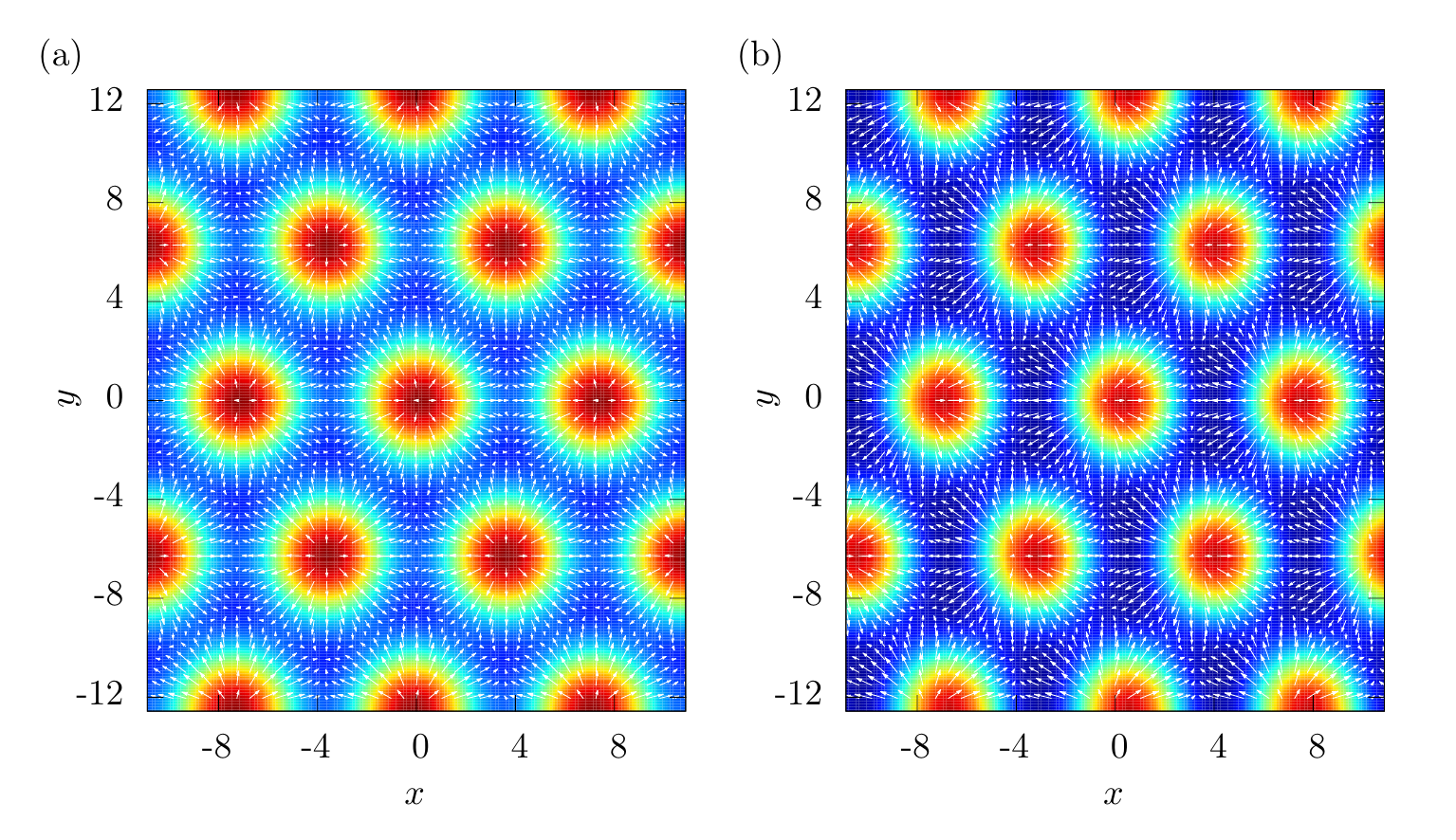}
	  \caption{Density $\psi(\mathbf{r})$ and polarization $\mathbf{P}(\mathbf{r})$ profiles in terms of a color map with overlaid white arrows, respectively, for (a) a resting ($v_0=0.25$) and (b) a traveling $(v_0=0.4>v_\mathrm{c})$ hexagonal pattern. In (a) the +1 defects of the polarization field coincide with the density maxima (and the net polarization is zero), while in (b) they are shifted with respect to one another, breaking the left-right symmetry. This shift generates a net polarization and results in net propulsion to the right with $c\approx0.3$. Parameters and domain size are as in Fig.~\ref{fig:cont_hex}.}
	 \label{fig:pol_hex}
\end{figure}

From time simulations in previous studies~\cite{MenzelLoewen, MenzelOhtaLoewenPhysRevE.89}, it is known that the activity parameter $v_0$ does not only lead to a transition from resting to traveling patterns, but also strongly influences the crystal structure. Here, we use numerical continuation to investigate if the different traveling patterns are connected via bifurcations and how the patterns are selected. 

Starting with a steady state hexagonal pattern at $v_0=0$ in a suitable domain, we follow the branch of hexagonal crystals in $v_0$. Figure~\ref{fig:cont_hex}(I) illustrates the chosen domain. Its aspect ratio corresponds to the ratio between the height $L_\mathrm{c}$ and the side length $L_\mathrm{a}$ of the equilateral triangles within the hexagon. The hexagons are oriented with one edge parallel to the direction of motion observed in time simulations close to the onset of motion (cf. Fig.~\ref{fig:DNS_ML}). Due to the employed boundary conditions (see Sec.~\ref{sec:num2d} for details) only motion along the $x$-axis is possible. 

The resulting bifurcation diagram is depicted in the main panel of Fig.~\ref{fig:cont_hex}. The branch of resting hexagons is shown in blue, whereas the traveling hexagonal pattern corresponds to the orange branch. At the critical activity $v_\mathrm{c}\approx0.3$, the resting pattern is destabilized in a drift-pitchfork bifurcation. The left bottom panel shows the characteristic growth of the drift velocity $c$; close to the drift instability $c\propto \sqrt{v_0-v_{\mathrm{c}}}$. The second small panel demonstrates that the quantity $||\psi||^2_2-||\mathbf{P}||^2_2$ crosses zero at the onset of motion. Note that the onset of motion at $v_\mathrm{c}\approx0.3$ corrects earlier studies~\cite{MenzelLoewen} and confirms the critical activity value found in \cite{PVWL2018pre}.

\begin{figure} \vspace*{0.cm}
\centering\hspace*{-0.5cm}
\includegraphics[width=0.5\textwidth]{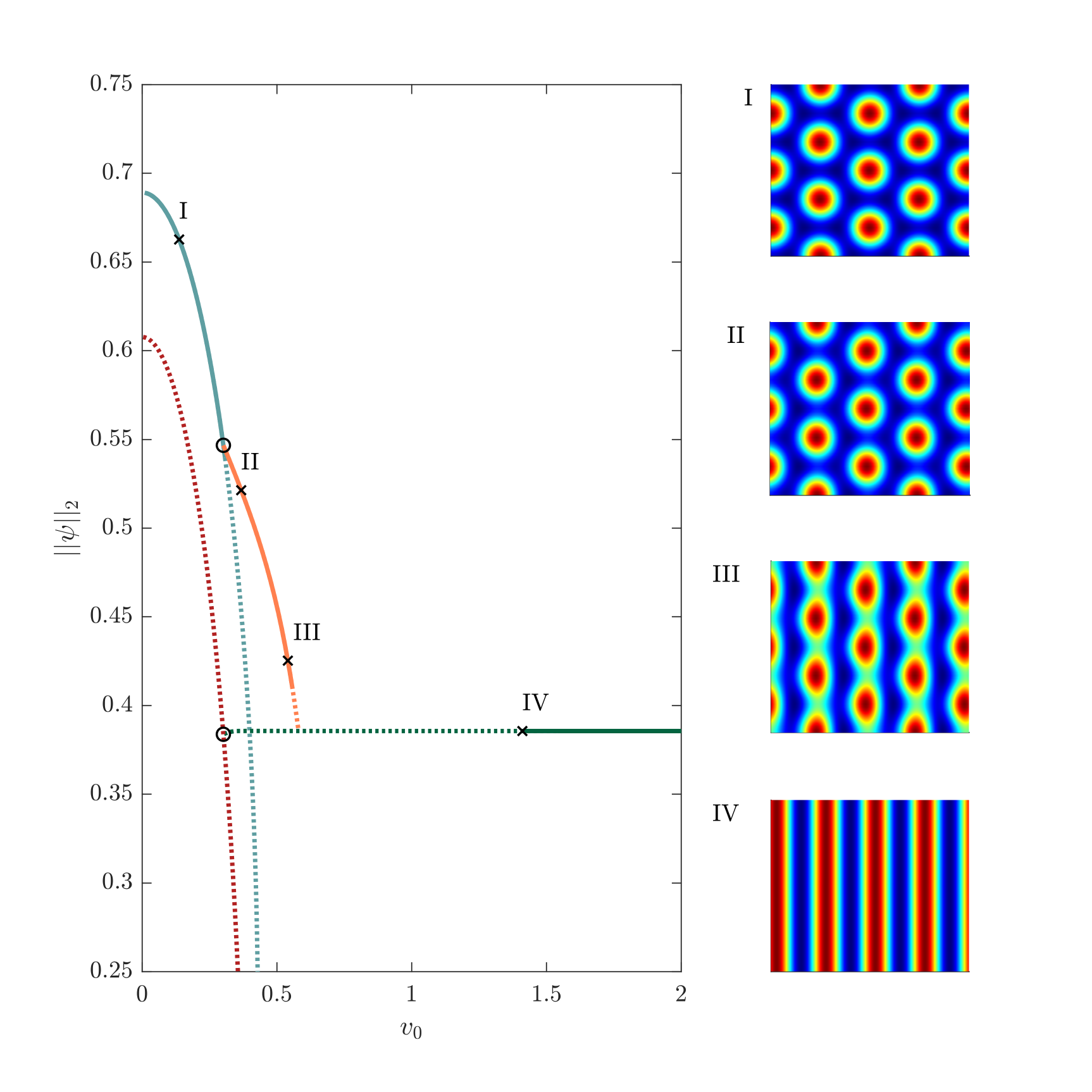}
\vspace*{-1cm}
\caption{
  The main panel shows the bifurcation diagram $||\psi||_2$ vs $v_0$ for hexagonal patterns for the onset of motion perpendicular to an edge. On the right selected solution profiles $\psi(\mathbf{r})$ at locations labeled I to IV in the bifurcation diagram are shown. Profiles II-IV travel in $x$-direction to the right. Resting hexagons (blue line) are stable (solid line) until a drift-pitchfork bifurcation at $v_\mathrm{c}\approx0.3$ where a branch of stable traveling hexagonal patterns (orange line) emerges. With increasing $v_0$, the traveling hexagonal pattern (e.g., profile~II) deforms into modulated stripes (e.g., profile~III). The branch terminates on the horizontal branch of traveling stripes (dark green line, e.g., profile~IV) that itself emerges in a drift-pitchfork bifurcation from an unstable branch of resting stripes (red line). The domain size is $V=4L_\mathrm{c}\times\,3L_\mathrm{a}$ and the remaining parameters are as in Fig.~\ref{fig:DNS_ML}.
}
	 \label{fig:cont_hex_stripes}
 \end{figure}

The four selected solution profiles $\psi(\mathbf{r})$ (I)-(IV) correspond to the locations indicated in the main panel. The density profiles illustrate how the hexagonal order of the crystal is preserved with increasing $v_0$. However, the individual density peaks change their shape from circular bumps towards oval and even rectangular peaks, cf. Fig.~\ref{fig:cont_hex}\,(III) and (IV). The branch of traveling hexagons is stable up to very high values of activity, in other words, we did not detect a destabilizing bifurcation on this branch.

Figure~\ref{fig:pol_hex} gives details on the symmetry breaking associated with the onset of motion of the hexagonal pattern. The density field $\psi(\mathbf{r})$ is given as a color map and the polarization field $\mathbf{P}(\mathbf{r})$ is indicated by white arrows. Panel (a) depicts the two fields in a resting crystal. As discussed for LS in Sec.~\ref{sec:v0} for resting states, the centers of the density peaks coincide with +1 defects of $\mathbf{P}$. One of the corresponding symmetries is broken beyond the onset of motion and the topological defects of $\mathbf{P}$ shift with respect to the peaks of $\psi$. Hence, when averaging the polarization over a density peak, a net polarization and drift emerge. Figure~\ref{fig:pol_hex}(b) shows a moving hexagonal crystal with a positive net polarization. In the red area of the maximum of $\psi$, more arrows point to the right than to the left and the crystal therefore moves to the right without change of shape. 

For the bifurcation diagram in Fig.~\ref{fig:cont_hex_stripes}, the orientation of the hexagon has been rotated by $90^{\circ}$, i.e., the drift is forced to occur perpendicular to the an edge of the hexagon. The domain size is adapted to match the hexagons by switching the lengths of $L_x$ and $L_y$ from Fig.~\ref{fig:cont_hex}. As in Fig.~\ref{fig:cont_hex}, the resting hexagonal pattern (blue branch) is destabilized in a drift-pitchfork bifurcation at $v_{\mathrm{c},\perp}=0.3015$ as compared to $v_{\mathrm{c},\parallel}=0.3008$ in Fig.~\ref{fig:cont_hex}. The slightly larger threshold is in agreement with results from time simulations where drift parallel to an edge (cf. Fig.~\ref{fig:cont_hex}) is found for motion at onset. Black circles highlight the drift bifurcations in the main panel of Fig.~\ref{fig:cont_hex_stripes}. Besides stable resting hexagons, an unstable resting stripe pattern exists in this setup (red branch). The resting stripes undergo a drift bifurcation at $v_\mathrm{c}\approx0.3$ as well.

In contrast to hexagons traveling parallel to an edge, the hexagons traveling perpendicular to an edge do not persist to arbitrarily high $v_0$ and instead terminate on a branch of traveling stripes (horizontal green line) in a supercritical pitchfork bifurcation. Along this branch, the crystal continuously changes from traveling deformed hexagons (Fig.~\ref{fig:cont_hex_stripes}, profile II) to traveling modulated stripes (III); moreover, the solutions lose stability in a Hopf bifurcation at $v_0\approx 0.6$ before reaching the termination point. The horizontal branch of moving stripes confirms the results for periodic states in 1D that also maintain a constant norm by shifting the relative positions between $\psi(x)$ and $P(x)$ with changing $v_0$. The traveling stripes eventually gain stability in a Hopf bifurcation at about $v_0=1.5$ after undergoing various bifurcations (not shown). At $v_0=1.5$, random initial conditions evolve into drifting stripes in numerical time stepping. Note that vertical stripes do not fit into the domain of Fig.~\ref{fig:cont_hex} as $L_x$ is not a multiple of $L_\mathrm{c}=2\pi$. In the parameter range where Fig.~\ref{fig:cont_hex_stripes} exhibits only unstable states, time simulations show either traveling rhombic patterns [cf.~Fig.~\ref{fig:DNS_ML}(c)] or states with a more intricate time dependence (not shown). 

\begin{figure} \vspace*{0.cm}
	 \centering
	 \includegraphics[width=0.5\textwidth]{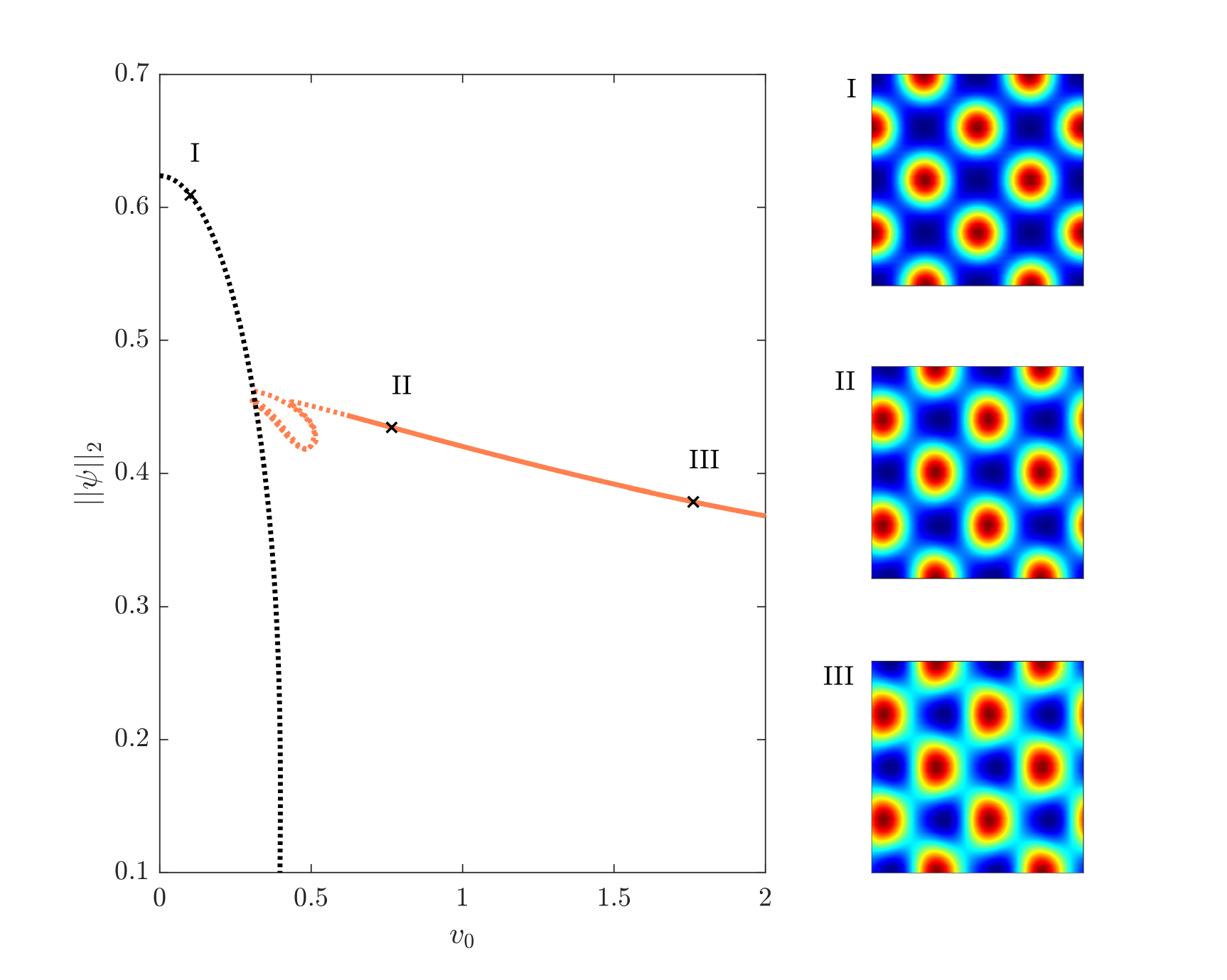}
	 \caption{
           The main panel shows the bifurcation diagram $||\psi||_2$ vs $v_0$ for square patterns oriented such that the motion is parallel to a diagonal of the square. On the right selected solution profiles $\psi(\mathbf{r})$ at locations labeled I to III in the bifurcation diagram are shown. Profiles II-III travel in the $x$-direction to the right. Resting squares (black line) are unstable. At drift-pitchfork bifurcations at $v_\mathrm{c}\approx0.3$ branches of unstable traveling square patterns (orange lines) emerge.  At $v_0\approx0.7$ traveling squares gain stability in a Hopf bifurcation. The domain size is $2\sqrt{2}L_\mathrm{c}\,\times\,2\sqrt{2}L_\mathrm{c}$ and the remaining parameters are as in Fig.~\ref{fig:DNS_ML}.
	}
	 \label{fig:cont_squares}
\end{figure}

Even though time simulations show a different direction at the onset of motion of the traveling hexagons, the detected branch of modulated stripes (Fig.~\ref{fig:cont_hex_stripes}, profile III) corresponds to a solution type that arises within large scale parameter scans presented in Sec.~\ref{sec:phasediagram_ML}. In addition, continuation confirms that resting stripes are unstable for all values of $v_0$ as suggested by time simulations. Since time simulations also point to rhombic patterns, we have also performed continuation on a square domain. Figure~\ref{fig:cont_squares} shows that the branch of squares traveling parallel to a diagonal (orange) becomes stable at about $v_0\approx0.7$, in perfect agreement with the traveling squares observed in time simulations [cf.~Fig.~\ref{fig:DNS_ML}\,(c)], and suggests that these stable traveling squares extend to arbitrarily large values of the activity parameter $v_0$. Squares traveling parallel to a side are expected as well, but were not computed.

\begin{figure}[t] 
	 \centering
	\includegraphics[width=0.45\textwidth]{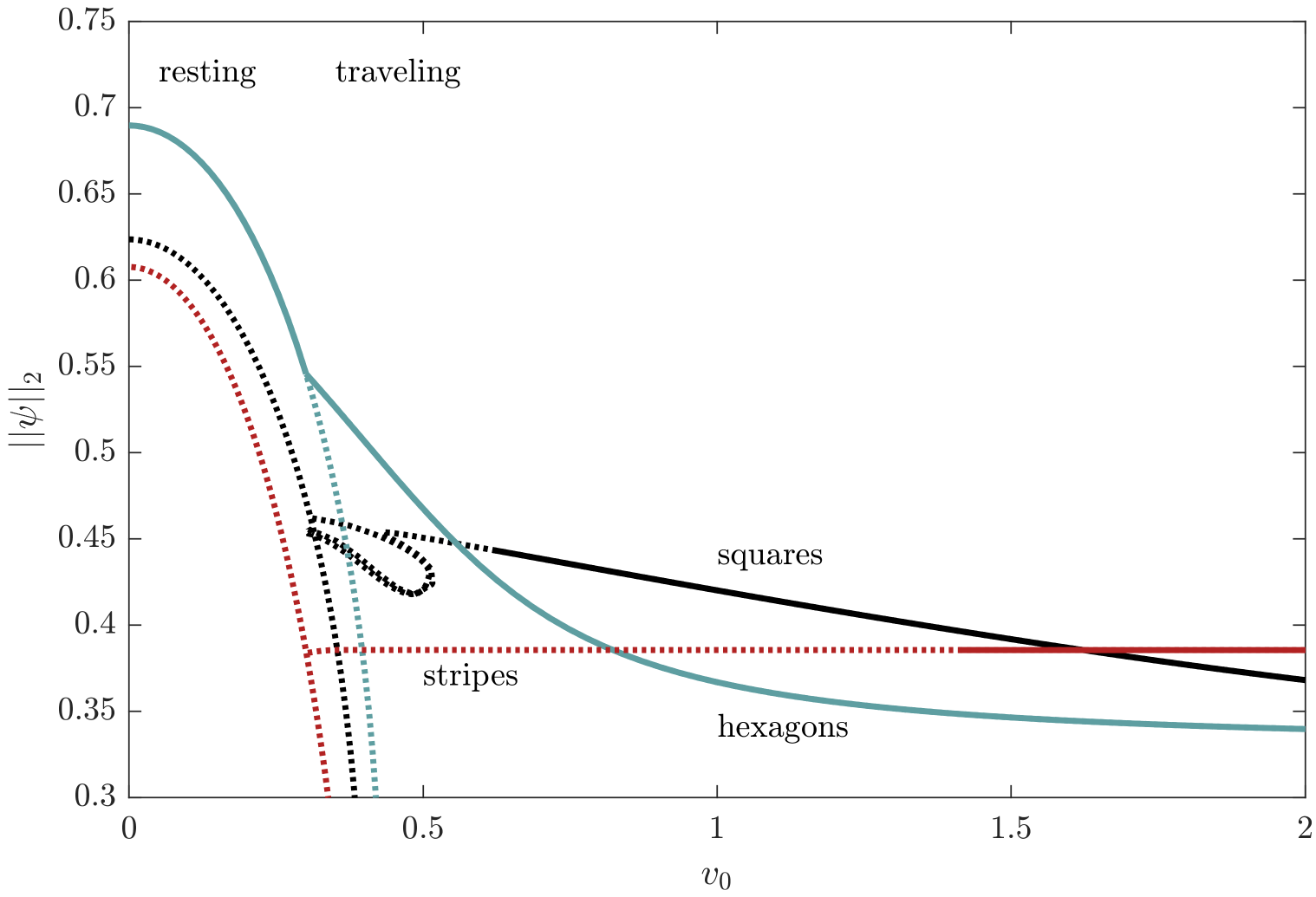}
	 \caption{Combined bifurcation diagrams for the resting and traveling periodic states in Figs.~\ref{fig:cont_hex}, \ref{fig:cont_hex_stripes} and~\ref{fig:cont_squares}. Note that the domain sizes differ between the different branches. All resting patterns undergo drift-pitchfork instabilities at nearly identical values of $v_0$, $v_{\mathrm{c}}\approx0.3$. Different traveling states coexist at large activity $v_0$.}
	 \label{fig:cont_all_crystals}
\end{figure}

Finally, Fig.~\ref{fig:cont_all_crystals} combines the results from continuation runs on different domains. Around practically identical values of $v_0$, $v_{\mathrm{c}}\approx 0.3$, all resting crystals undergo drift instabilities. As for the resting states at small $v_0$, only hexagons (blue branch) are stable (solid line). Traveling squares and traveling stripes gain stability at higher values of $v_0$ that are in perfect agreement with numerical time simulations. Numerical continuation suggests that different traveling crystals coexist. In the simulations the domain and in particular its aspect ratio appear to select the moving pattern.

  %

\subsection{Morphological phase diagram}
\label{sec:phasediagram_ML}

\begin{figure}	
	\centering	
%
\vspace*{-0cm}\begin{overpic}[width=0.5\textwidth]{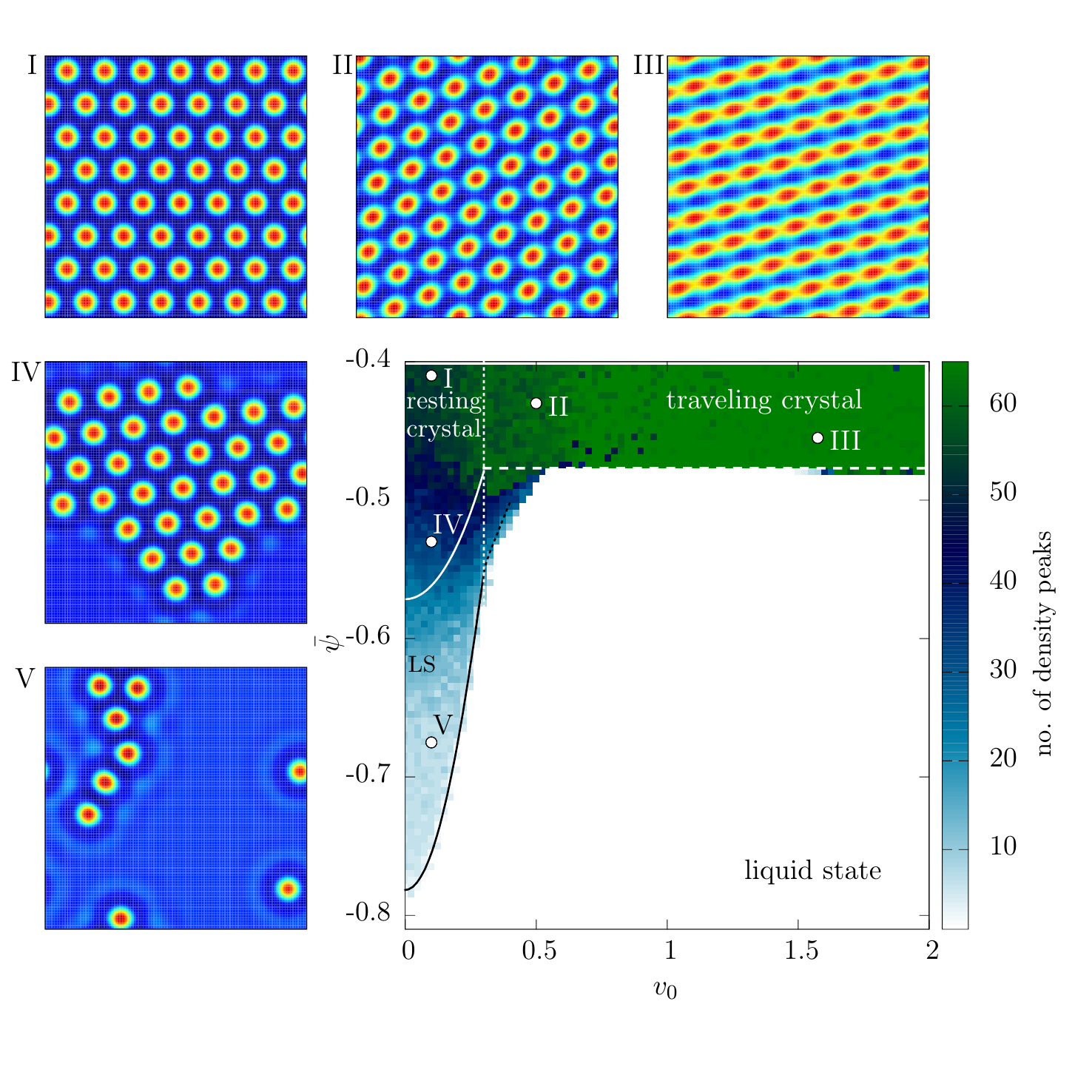}
	\put(47.3,79.7){\rotatebox{-65.}{\makebox(0,0){\strut{}\textcolor{white}{\normalsize{$\Longrightarrow$}}}}}%
	\put(68.5,84.2){\rotatebox{-76.}{\makebox(0,0){\strut{}\textcolor{white}{\normalsize{$\Longrightarrow$}}}}}%

	\end{overpic} 
	 \vspace*{-1.3cm}
	 \caption{Morphological phase diagram for the aPFC model accompanied by selected density profiles at locations labeled I-V obtained from systematic simulations. Large panel: Different states are characterized by the total number of density peaks that form in a rectangular domain of size $7L_\mathrm{a} \times 8L_\mathrm{c}$ as indicated by color coding.
	The various lines in the diagram, the initial conditions of the simulations, and the peak counting procedure are described in the text. The liquid state refers to a uniform density phase with zero peaks (white area). LS exist in the regions marked in blue. Periodic hexagonal patterns (green) fill the domain with 56 density peaks (I). Around $v_0>1$ the number of peaks slightly increases as resting hexagonal patterns (I) begin transforming towards traveling rhombic patterns (II) with a smaller wavelength allowing for more density peaks. Arrows indicate the direction of motion. At even higher $v_0$, traveling stripe patterns (III) arise. These remain spatially modulated so that individual peaks can still be located on each ridge and the number of density peaks does not drop. (IV) and (V) give examples of resting LS coexisting with the liquid phase. The remaining parameters are $\epsilon=-0.98$, $C_1=0.2$, $C_2=0$ and $D_\mathrm{r}=0.5$ as in~\cite{MenzelLoewen}.}
	 \label{fig:phasediagram_profiles_ML}
	 \end{figure}

In order to complete the picture of crystalline states and the influence of $v_0$ and $\bar{\psi}$, we again perform numerous time simulations and construct a morphological phase diagram in the $(v_0,\bar{\psi})$-plane for the parameter set employed in this section. The time simulations are carried out in the same way as previously described in Sec.~\ref{sec:phasediagram} and again LS and various periodic states are distinguished in terms of the number of peaks of $\psi(\mathbf{r})$. The resulting phase diagram is shown in the main panel of Fig.~\ref{fig:phasediagram_profiles_ML}. 

The parameter set used is from Ref.~\cite{MenzelLoewen} and includes a high value of the diffusion constant, $C_1=0.2$ -- twice the value used in Sec.~\ref{sec:v0}. The high diffusion leads to, first, a higher $v_{\mathrm{c}}\approx0.3$ (cf. vertical dotted line) and, second, it suppresses the existence of LS for increasing activity (no blue areas for $v_0>0.5$). The green crystalline area exhibits density fields with more than 56 peaks for high activities. Here, solution profiles show traveling rhombic patterns (Fig.~\ref{fig:phasediagram_profiles_ML}, profile II) with a smaller wavelength than the hexagonal states. Thus, more peaks fit into the considered domain. At higher mean densities towards $\bar{\psi}=-0.4$ the rhombic patterns transform into a stripe pattern. However, the stripes are still sufficiently modulated in space to account for a high number of peaks as depicted in Fig.~\ref{fig:phasediagram_profiles_ML}, profile III. Accordingly, the number of counted density peaks does not decrease.

Figure~\ref{fig:phasediagram_profiles_ML} gives a detailed overview of the active crystals and crystallites that arise for the parameter set used in Ref.~\cite{MenzelLoewen}. It also evidences a neat interplay between numerical continuation and time simulations as the results from fold continuation (black lines) nicely bound the region of existence of LS. 

\section{\label{sec:summary}Summary and conclusions}

We have studied in considerable depth the bifurcation structure of an active phase-field-crystal model in two spatial dimensions. This model, first introduced in Ref.~\cite{MenzelLoewen}, describes a variety of resting and traveling spatially extended and spatially localized structures.  

First, using the mean concentration $\bar\psi$ as the control parameter, we have analyzed how the classical slanted snakes-and-ladders structure (slanted homoclinic snaking) known from the phase-field-crystal model \cite{TARG2013pre} is modified by activity. In particular, we have shown that with increasing activity, one finds a critical value for the onset of motion of both domain-filling crystals and the various localized states associated with them. In general, an increase in activity suppresses resting localized and crystalline states. Resting LS ultimately annihilate in saddle-node bifurcations at critical values of the activity parameter that are similar for all the states studied, while resting periodic or crystalline states disappear in supercritical pitchfork bifurcation of the homogeneous or liquid state. In other words, activity eventually melts all resting crystalline structures as the driving force overcomes the attractive forces that stabilize the equilibrium crystals and the crystallites that exist in the reference system without activity. This melting of equilibrium clusters due to activity has been observed in Brownian dynamics simulations of Ref.~\cite{ReBH2013pre} for self-propelled particles with short-range attraction.

However, at values of the activity below this melting point, the branches of resting states exhibit drift bifurcations for suitable diffusion and mean densities, generating branches of traveling states. These may exist stably within certain ranges of activity as shown here by numerical two-parameter continuation of the relevant bifurcations. In other words, although activity may melt traveling crystallites, there are extended parameter regimes where this is not the case. In fact, we have found that while high activity melts most traveling localized states, i.e., traveling crystalline patches, this is not the case for traveling periodic states, i.e., traveling domain-filling crystals. These can be driven with arbitrarily high activity and then exhibit correspondingly high drift velocities. We believe that this is most likely the case because the periodicity of the domain-filling crystals is fixed, while traveling localized states naturally adapt their peak to peak spacing to the imposed parameter values. This additional degree of freedom may make such states less stable. Note that the crystallites we have found are not related to the motility-induced clusters discussed, e.g., in \cite{Ginot2015prx,SSWK2015prl,CaTa2015arcmp}. The latter effect has not yet been found in an active PFC model since such models are generally geared to studies of how equilibrium crystallization is modified by nonzero activity. Whether such models may also describe motility-induced clustering, especially when allowing for spontaneous polarization ($C_1<0$, $C_2>0$), is a question of some importance.

Next, we have investigated the region of existence of traveling localized states and showed that such TLS are generic solutions in extended regions of the plane spanned by the mean concentration and activity. While broader TLS with three and more peaks quickly vanish into the homogeneous background, narrow localized states (with one and two density peaks) can be driven at quite high activities where they reach high velocities. This does not seem to be the case in the nonvariational systems studied in \cite{HoKn2011pre,BuDa2012sjads}. Thus a future comparative study of the present system, the systems studied in \cite{HoKn2011pre,BuDa2012sjads} and those reviewed and discussed in \cite{KoTl2007c} would be beneficial.

A substantial focus of the paper has been on the nature of the onset of motion of the competing localized and extended structures. We found that this occurs at critical values of the activity that depend only weakly on the size of a particular localized state or the number of density peaks within it. We have shown that a previously derived criterion for the onset of motion of active crystals in 1D also holds in two dimensions, namely, that the zero crossing of the difference of the squared norms of the two steady fields ($||\psi_{0}||_2^2-||P_{0}||_2^2$) marks the onset of motion for all localized and extended crystalline states. This criterion holds at the drift-pitchfork bifurcation of $\kappa$-symmetric states and may be used to determine the critical strength of the activity parameter that is needed for collective traveling motion. It also determines the onset of drift of asymmetric states via the drift-transcritical bifurcation. Whether such simple criteria can be derived for more complicated active matter models that capture faithfully the specific properties of laboratory systems and the active particles at hand remains to be investigated.

The onset of motion in the aPFC model studied here differs from that in the nonvariational Swift-Hohenberg equations studied in \cite{HoKn2011pre}. There, at any value of the driving parameter in front of the nonvariational term, all asymmetric states drift and all symmetric states are at rest. Here, however, the special form of the coupling of the two fields allows for resting asymmetric states even at a finite activity parameter, a nongeneric feature of the model that will be investigated further in future work. 
Within the aPFC model both symmetric and asymmetric states undergo sharp transitions to drift as the activity parameter $v_0$ increases. However, only the former are expected to be present in generic $\kappa$-symmetric models, with the latter replaced by continuous or imperfect transitions. Because of this the results of the present work are expected to assist greatly in the computations of drifting states in such models, particularly those lying on disconnected branches associated with such imperfect bifurcations. This topic will also be the subject of a future study.

The additional degrees of freedom present in 2D lead to considerably more complex bifurcation diagrams than in 1D \cite{OphausPRE18} largely because more states are possible and the fact that these states can drift in more than one direction. Besides translation modes, rotational modes can also be destabilized and the particular direction of the drift with respect to symmetry axes of the LS has to be taken into account. A rotationally symmetric one-peak LS has shown many similarities to the 1D case, while the less symmetric dumbbell-shaped two-peak LS turned out to be unstable for any nonvanishing value of activity. For the employed value of the effective temperature, the active crystallites exhibit slanted homoclinic snaking and resting LS as well as traveling LS exhibit similar behavior. The LS, whether resting or traveling, gradually grow in size as one follows the LS branch until they fill the entire domain and respectively terminate on a periodic resting or traveling solution. At lower values of the effective temperature the snaking behavior apparently ceases and is replaced by new behavior the details of which remain unclear. 

In 2D, the activity parameter also strongly influences the crystal structure of space-filling, fully periodic solutions. We have identified a multistability region with stable traveling hexagons, traveling rhombuses and traveling stripes. Here, finite size effects such as the aspect ratio of the domain control pattern selection but it is evident than in the thermodynamic limit the phase diagram must be highly complex. The morphological phase diagrams we have constructed to compress information from time simulations and numerical continuation provide an indication of this complexity. The figure demonstrates the impressive capability of fold continuation to predict limits of existence of various solutions found in time simulations.

Finally, we highlight a number of questions that merit further investigation. Our main aim has been to establish an overview of the rather involved overall bifurcation structure that is related to the onset of motion in continuum models of active crystals. As experimental studies often focus on the collective behavior of many interacting particles and clusters \cite{theurkauff2012prl,Ginot2015prx,ginot2018aggregation}, we need to investigate further whether it is possible to derive statistical models from single cluster bifurcation studies such as the present one. Such a methodology has recently been presented for ensembles of sliding drops \cite{WTEG2017prl}. The aPFC model studied here involves a rather simple coupling of concentration and polarization and excludes spontaneous polarization. These limitations are responsible for the presence of the $\kappa$-symmetry of the model that is in turn responsible for the presence of drift bifurcations that govern so much of the behavior reported here. It is necessary therefore that the results obtained here regarding the onset of motion should be compared to related results regarding the bifurcation structure of other, more realistic models of active matter. This will allow one to develop a clearer general understanding of the observed multistability of states and associated hysteresis effects as well as of the thresholds for qualitative changes in behavior. The present study may serve as a road map for such analyses.

\section*{Acknowledgements}

We acknowledge support through the doctoral school ``Active living fluids'' funded by the German French University  (Grant No. CDFA-01-14). LO wishes to thank ``Studienstiftung des deutschen Volkes'' and ``IP@WWU'' for financial support, and Fenna Stegemerten, Max Holl and Tobias Frohoff-H\"ulsmann for frequent discussions. A special thanks goes to Tobias Frohoff-H\"ulsmann for his pde2path support and to Hannes Uecker for incorporating new features into pde2path. The work of EK was supported in part by the National Science Foundation under grant DMS-1908891.


%
%
\end{document}